\documentclass[12pt]{iopart}

\usepackage{iopams}
\usepackage{amsthm}
\usepackage{graphicx}
\usepackage{setstack}
\usepackage{xcolor}
\usepackage{ulem}

\newtheorem{Proposition}{Proposition}
\newtheorem{Lemma}[Proposition]{Lemma}

\newcommand{\bmath}[1]{\mbox{{\boldmath{{$#1$}}}}}
\newcommand{\drs}{\partial _{r_{*}}}
\newcommand{\Diag}{\mathrm {Diag}}

\eqnobysec

\begin{document}

\title[Stability of dyons in EYM]{On the stability of dyons and dyonic black holes in Einstein-Yang-Mills theory}

\author{Brien C. Nolan${}^{1}$ and Elizabeth Winstanley${}^{2}$}

\address{${}^{1}$
School of Mathematical Sciences,
Dublin City University,
Glasnevin,
Dublin 9,
Ireland}

\address{${}^{2}$
Consortium for Fundamental Physics,
School of Mathematics and Statistics, \\
The University of Sheffield,
Hicks Building,
Hounsfield Road,
Sheffield.
S3 7RH
United Kingdom}

\eads{\mailto{brien.nolan@dcu.ie}, \mailto{E.Winstanley@sheffield.ac.uk}}

\begin{abstract}
We investigate the stability of four-dimensional dyonic soliton and black hole solutions of ${\mathfrak {su}}(2)$ Einstein-Yang-Mills theory in anti-de Sitter space.
We prove that, in a neighbourhood of the embedded trivial (Schwarzschild-)anti-de Sitter solution, there exist non-trivial dyonic soliton and black hole solutions of the field equations
which are stable under linear, spherically symmetric, perturbations of the metric and non-Abelian gauge field.
\end{abstract}


\section{Introduction}
\label{sec:intro}

Since the discovery of non-trivial soliton \cite{Bartnik} and black hole \cite{AFBHs} solutions of the four-dimensional ${\mathfrak {su}}(2)$ Einstein-Yang-Mills (EYM) equations in asymptotically flat space-time, the EYM system has been studied extensively (see \cite{Volkov1} for a review).
For ${\mathfrak {su}}(2)$ gauge group, the gauge field of non-trivial solutions in four-dimensional asymptotically flat space-time is purely magnetic \cite{Ershov} and, furthermore, these solutions are unstable under linear, spherically symmetric, perturbations \cite{Straumann}.

The properties of EYM solutions in asymptotically anti-de Sitter (adS) space-time are very different from those in asymptotically flat space-time.
The first difference is the existence of four-dimensional, spherically symmetric,
purely magnetic soliton \cite{Bjoraker} and black hole \cite{Winstanley1} solutions
of ${\mathfrak {su}}(2)$ EYM in adS which are stable under linear, spherically symmetric, perturbations.
Subsequently it was shown that there exist both soliton and black hole solutions which are stable under general linear perturbations of the metric and gauge field \cite{Sarbach}.
If the gauge group is enlarged to ${\mathfrak {su}}(N)$, purely magnetic, spherically symmetric, soliton and black hole solutions with $N-1$ gauge field degrees of freedom exist \cite{Baxter1}.
It can be proven that at least some of these are stable under linear, spherically symmetric perturbations \cite{Baxter}.

The second surprising feature of solutions of ${\mathfrak {su}}(2)$ EYM in adS is the existence of non-trivial, spherically symmetric, dyonic solitons and black holes \cite{Bjoraker,Nolan}.
For these solutions the gauge field has a non-trivial electric part as well as a magnetic part.
Properties of these spherically symmetric dyonic solitons and black holes were explored numerically in \cite{Bjoraker}.
The existence of non-trivial dyonic solutions in a neighbourhood of the trivial (embedded Schwarzschild-adS) solution was proven in \cite{Nolan}.
Although these dyonic solutions were discovered numerically over fifteen years ago, their stability has remained an open question which we address in this paper.

We consider static, spherically symmetric, dyonic soliton and black hole solutions of ${\mathfrak {su}}(2)$ EYM in adS.
In section \ref{sec:dyons} we introduce the action and field equations, and briefly review some of the properties of the static equilibrium solutions
\cite{Bjoraker,Nolan}.  Next, in section \ref{sec:perteqns}, we derive the equations governing time-dependent, linear, spherically symmetric perturbations of the static equilibrium solutions.
The analysis results in a pair of coupled Schr\"odinger-like equations for two of the perturbations. The third independent perturbation is governed by a constraint
equation which does not involve any derivatives with respect to time.
Section \ref{sec:stable} contains our proof of the existence of non-trivial dyonic solitons and black holes, in a neighbourhood of the embedded trivial solution, which are stable under the linear perturbations.
Finally we present our conclusions in section \ref{sec:conc}.

\section{Dyons and dyonic black holes in ${\mathfrak {su}}(2)$ Einstein-Yang-Mills theory}
\label{sec:dyons}

In this section we introduce the action and field equations for ${\mathfrak {su}}(2)$ Einstein-Yang-Mills theory with a negative cosmological constant.
We also briefly review some of the properties of the static, spherically symmetric, dyon and dyonic black hole solutions of this theory, which were found numerically in \cite{Bjoraker} and whose existence was proven in \cite{Nolan}.

\subsection{Ansatz and field equations}
\label{sec:ansatz}

We begin with the action for Einstein-Yang-Mills theory in four-dimensional space-time with a cosmological constant $\Lambda $:
\begin{equation}
S_{\mathrm {EYM}} = \frac {1}{2} \int d ^{4}x {\sqrt {-g}} \left[
R - 2\Lambda
- \Tr \, F_{\tau \nu }F ^{\tau \nu }
\right] ,
\label{eq:action}
\end{equation}
where $R$ is the Ricci scalar, $g$ is the metric determinant and $F_{\tau \nu }$ is the Yang-Mills gauge field. $\Tr $ denotes a Lie algebra trace.
Here and throughout this paper, the space-time has signature $(-,  + , + , +)$, we use units in which $4\pi G = 1= c$ and we have fixed the gauge coupling constant to be equal to unity.
We consider a negative cosmological constant $\Lambda <0$ and the gauge Lie algebra is ${\mathfrak {su}}(2)$.

Varying the action (\ref{eq:action}) we obtain the field equations
\numparts
\begin{eqnarray}
2T_{\tau \nu } & = &  R_{\tau \nu } - \frac {1}{2} R g_{\tau \nu } +
\Lambda g_{\tau \nu },
\label{eq:Einstein} \\
0 &  = &  D_{\tau } F_{\nu }{}^{\tau } = \nabla _{\tau } F_{\nu }{}^{\tau }
+ \left[ A_{\tau }, F_{\nu }{}^{\tau } \right] ,
\label{eq:YM}
\end{eqnarray}
\endnumparts
where the Yang-Mills field strength tensor takes the form
\begin{equation}
F_{\tau \nu } = \partial _{\tau }A_{\nu } - \partial _{\nu }A_{\tau } +
\left[ A_{\tau },A_{\nu } \right] ,
\label{eq:Fmunu}
\end{equation}
with $A_{\tau }$ the Yang-Mills gauge potential, and $\left[ A_{\tau }, A_{\nu } \right] $ denoting the Lie algebra commutator.
The  stress energy tensor is
\begin{equation}
T_{\tau \nu } =
\Tr F_{\tau \lambda } F_{\nu }{}^{\lambda }
- \frac {1}{4} g_{\tau \nu }
\Tr F_{\lambda \rho} F^{\lambda \rho } .
\label{eq:Tmunu}
\end{equation}

In this paper we are interested in the stability of static, spherically symmetric, dyon and dyonic black hole solutions of the field equations (\ref{eq:Einstein}--\ref{eq:YM}).  In the next section we shall consider time-dependent, linear, spherically symmetric perturbations of the static equilibrium
solutions, so we consider a time-dependent, spherically symmetric metric as follows
\begin{equation}
ds^{2} = -\mu (t,r) S(t,r)^{2} \, dt^{2} + \mu (t,r)^{-1} \, dr^{2} + r^{2} \left( d\theta ^{2} + \sin ^{2} \theta \,  d\phi ^{2} \right) ,
\label{eq:metric}
\end{equation}
where the metric functions $\mu (t,r)$ and $S(t,r)$ depend on time $t$ and the radial co-ordinate $r$.
We may write the metric function $\mu (t,r)$ in the alternative form
\begin{equation}
\mu (t,r) = 1 - \frac {2m(t,r)}{r} + \frac {r^{2}}{\ell ^{2}},
\label{eq:mu}
\end{equation}
where the adS radius of curvature $\ell $ is given by
\begin{equation}
\ell ^{2} = -\frac {3}{\Lambda }.
\end{equation}

The time-dependent, spherically symmetric ${\mathfrak {su}}(2)$ Yang-Mills gauge potential $A_{\tau }$ can be written as follows, after an appropriate choice of gauge \cite{Kunzle}:
\begin{equation}
A = {\mathcal {A}} \, dt + {\mathcal {B}} \, dr + \frac {1}{2} \left( C - C^{H} \right)  d \theta
- \frac {i}{2} \left[ \left( C + C^{H} \right) \sin \theta + D\cos \theta \right]  d \varphi ,
\label{eq:gauge}
\end{equation}
where ${\mathcal {A}}$,  ${\mathcal {B}}$, $C$ and $D$ are $2\times 2$ matrices, given by
\begin{eqnarray}
{\mathcal {A}}
& = & \frac {i}{2} \left(
\begin{array}{cc}
\alpha (t,r) & 0 \\
0 & -\alpha (t,r)
\end{array}
\right) ,
\qquad
{\mathcal {B}}
= \frac {i}{2} \left(
\begin{array}{cc}
\beta (t,r) & 0 \\
0 & -\beta (t,r)
\end{array}
\right) ,
\nonumber
\\
C  & = & \left(
\begin{array}{cc}
0 & \omega (t,r)e^{i\gamma (t,r)} \\
0 & 0
\end{array}
\right) ,
\qquad
D = \left(
\begin{array}{cc}
1 & 0 \\
0 & -1
\end{array}
\right) .
\end{eqnarray}
Here, $\alpha (t,r)$, $\beta (t,r)$, $\gamma (t,r)$ and $\omega (t,r)$ are real functions of time $t$ and the radial co-ordinate $r$.  The matrix $C^{H}$
is the Hermitian conjugate of the matrix $C$.

\subsection{Static, spherically symmetric, dyons and dyonic black holes}
\label{sec:static}

For static equilibrium solutions of the field equations, the metric functions $m=m_{0}(r)$ and $S=S_{0}(r)$ now depend only on the radial co-ordinate $r$.
By a choice of gauge \cite{Kunzle}, the gauge field function $\beta $ can be set to zero, and then one of the Yang-Mills equations reduces to $\gamma =0$.
The remaining gauge field functions, $\alpha =\alpha _{0}(r)$ and $\omega =\omega _{0}(r)$, are also functions of $r$ only.

The field equations (\ref{eq:Einstein}, \ref{eq:YM}) then reduce to the following static field equations:
\numparts
\begin{eqnarray}
m_{0} ' & = &
\frac {r^{2}\alpha _{0}'^{2}}{2S_{0}^{2}} + \frac {\alpha _{0}^{2}\omega _{0}^{2}}{\mu _{0}S_{0}^{2}}
+ \mu _{0} \omega _{0}'^{2} + \frac {1}{2r^{2}} \left( 1- \omega _{0}^{2}\right) ^{2} ,
\label{eq:static1}
\\
\frac {S_{0}'}{S_{0}} & = &
\frac {2\omega _{0}'^{2}}{r} + \frac {2\alpha _{0}^{2}\omega _{0}^{2} }{r\mu _{0}^{2} S_{0}^{2}} ,
\label{eq:static2}
\\
0 & = &
\mu _{0}\alpha _{0} '' + \left( \frac {2\mu _{0}}{r} - \frac {\mu _{0}S_{0}'}{S_{0}} \right) \alpha _{0}' - \frac {2\alpha _{0}\omega _{0}^{2}}{r^{2}} ,
\label{eq:static3}
\\
0 & = &
\mu _{0}\omega _{0}'' + \left( \mu _{0}' + \frac {\mu _{0}S_{0}'}{S_{0}} \right) \omega _{0}' + \frac {\omega _{0}}{r^{2}} \left( 1-\omega _{0}^{2}\right)
+ \frac {\alpha _{0}^{2}\omega _{0} }{\mu _{0}S_{0}^{2}} ,
\label{eq:static4}
\end{eqnarray}
\endnumparts
where a prime $'$ denotes differentiation with respect to $r$.
The static field equations (\ref{eq:static1}--\ref{eq:static4}) possess the following symmetries. Firstly they are invariant under the transformation
$\alpha _{0} \rightarrow -\alpha _{0}$; secondly the transformation $\omega _{0} \rightarrow -\omega _{0}$ also leaves them unchanged;  and finally
they are preserved by the scaling symmetry:
\begin{equation}
S_{0} \rightarrow \lambda S_{0}, \qquad \alpha _{0} \rightarrow \lambda \alpha _{0},
\label{eq:scaling}
\end{equation}
for any constant $\lambda $. The scaling symmetry (\ref{eq:scaling}) arises  due to the invariance of the static metric and gauge potential under rescalings of the time co-ordinate $t\rightarrow \lambda ^{-1}t$.
When the metric and gauge potential are time-dependent, the gauge freedom remaining in rescaling the time co-ordinate is discussed in section \ref{sec:gauge}.

The static field equations (\ref{eq:static1}--\ref{eq:static4}) have three singular points of interest. These are located at the origin $r=0$ (relevant only for soliton solutions), at the black hole horizon (corresponding to zeros $r=r_h$ of the metric function $\mu$, if there are any), and as $r\rightarrow \infty $. As pointed out in \cite{Nolan}, while zeros of the metric function $S$ yield a fourth possible singular point, these are not of relevance to the classes of solutions considered here. Suitable boundary conditions therefore have to be imposed on the field variables at the singular points $r=0, r=r_h$ and $r\to\infty$.
Near the origin, the field variables take the form \cite{Bjoraker,Nolan}
\begin{eqnarray}
m_{0}(r)  =  \left( \frac {\alpha _{1}^{2}}{2S_{1}^{2}} + 2\omega _{2}^{2} \right)
r^{3}+O( r^{4} ),
\nonumber \\
S_{0}(r) =  S_{1} + \left( \frac {\alpha _{1}^{2}}{S_{1}} + 4S_{1} \omega _{2}^{2} \right) r^{2} + O( r^{3}) ,
\nonumber \\
\alpha_{0} (r)  =  \alpha _{1} r + \frac {\alpha _{1}}{5} \left(
\frac {2\alpha _{1}^{2}}{S_{1}^{2}} + 8 \omega _{2}^{2} + 2\omega _{2}
- \frac {1}{\ell ^{2}}
\right) r^{3}
+ O( r^{4} ) ,
\nonumber \\
\omega_{0} (r)  =  1 +\omega _{2} r^{2}+ O( r^{3} ) ,
\label{eq:origin}
\end{eqnarray}
and the solutions are parameterized by the constants $S_{1}$, $\alpha _{1}$, $\omega _{2}$ and the adS radius of curvature $\ell $.
In a neighbourhood of the black hole event horizon, the corresponding expansion of the field variables is \cite{Bjoraker,Nolan}
\begin{eqnarray}
m_{0} (r)  =
\frac {r_{h}}{2} + \frac {r_{h}^{3}}{2\ell ^{2}} + m'_{h} \left( r - r_{h} \right)
+ O(r-r_{h})^{2},
\nonumber
\\
S_{0} (r)  =
S_{h} + S'_{h} \left( r- r_{h} \right) + O(r-r_{h})^{2},
\nonumber
\\
\alpha_{0} (r)  =
\alpha '_{h} \left( r- r_{h} \right) + O(r-r_{h})^{2},
\nonumber
\\
\omega_{0} (r)  =
\omega _{h}+ \omega '_{h} \left( r - r_{h} \right) + O(r-r_{h})^{2},
\label{eq:horizon}
\end{eqnarray}
where $m_{h}'$, $S_{h}'$ and $\omega _{h}'$ are given by the field equations
(\ref{eq:static1}--\ref{eq:static4})
in terms of $S_{h}$, $\alpha _{h}'$ and $\omega _{h}$:
\begin{eqnarray}
m_{h}'  =  \frac {r_{h}^{2}\alpha _{h}'^{2}}{2S_{h}^{2}}+
\frac {\left( 1-\omega _{h}^{2}\right) ^{2}}{2r_{h}^{2}} ,
\nonumber
\\
S_{h}'  =
\frac {2\alpha _{h}'^{2}\omega _{h}^{2}}{S_{h}r_{h}\mu '(r_{h})^{2}}
+ \frac {2S_{h}\omega _{h}'^{2}}{r_{h}} ,
\nonumber
\\
\omega _{h}'  =
\frac {\omega _{h}}{r_{h}^{2} \mu '(r_{h})} \left( \omega _{h}^{2} -1 \right) ,
\end{eqnarray}
where
\begin{equation}
\mu '(r_{h}) = \frac {1}{r_{h}} - \frac {2m_{h}'}{r_{h}} + \frac {3r_{h}}{\ell ^{2}} >0 ,
\end{equation}
so that the constants $S_{h}$, $\alpha _{h}$ and $\omega _{h}$, together with the adS radius of curvature $\ell $, parameterize the solutions.
At infinity the field variables have the following behaviour \cite{Bjoraker,Nolan}:
\begin{eqnarray}
m_{0}(r)  =  M - \frac {1}{r} \left[
\frac {d_{1}^{2}}{2} + \alpha _{\infty }^{2} \omega _{\infty }^{2} \ell ^{2}
+ \frac {c_{1}^{2}}{\ell ^{2}}
+ \frac {\left( 1- \omega _{\infty }^{2} \right) ^{2}}{2}
\right] + O(r^{-2}) ,
\nonumber
\\
S_{0}(r)  =  1 - \frac {1}{2r^{4}} \left(
\alpha _{\infty }^{2} \omega _{\infty }^{2} \ell ^{4} + c_{1}^{2}
\right)  + O(r^{-5}) ,
\nonumber
\\
\alpha _{0}(r)  =  \alpha _{\infty } + \frac {d_{1}}{r} + O(r^{-2}),
\nonumber
\\
\omega _{0}(r)  =  \omega _{\infty } + \frac {c_{1}}{r} + O(r^{-2}),
\label{eq:infinity}
\end{eqnarray}
where $M$, $\alpha _{\infty }$, $\omega _{\infty }$, $c_{1}$ and $d_{1}$ are arbitrary constants.
The fact that $S_{0}(r) \rightarrow 1$ as $r\rightarrow \infty $ fixes the parameters $S_{1}$ and $S_{h}$ in the expansions of $S_{0}$ near the origin (\ref{eq:origin}) and event horizon (\ref{eq:horizon}), respectively.
In practice, however, we can regard $S_{1}$ and $S_{h}$ as free parameters, since, if $S_{\infty }\neq 1$, a scaling transformation (\ref{eq:scaling}) with $\lambda = S_{\infty }^{-1}$ can always be applied.

The field equations (\ref{eq:static1}--\ref{eq:static4}) possess a trivial solution given by
\begin{equation}
\alpha _{0}(r) \equiv 0, \qquad \omega _{0}(r) \equiv \pm 1, \qquad m_{0}(r) \equiv M, \qquad S_{0}(r) \equiv 1.
\label{eq:SadS}
\end{equation}
For $M>0$ this is the Schwarzschild-adS black hole; for $M=0$ this is pure adS space-time.
There are also embedded (electrically and magnetically charged) Abelian Reissner-Nordstr\"om-adS solutions of the static field equations \cite{Nolan}, but we shall not consider these further in this paper.
Purely magnetic solutions, whose properties are discussed in \cite{Bjoraker,Winstanley1,Winstanley}, arise on setting $\alpha _{0}\equiv 0$.

\begin{figure}
\begin{center}
\includegraphics[width=10cm]{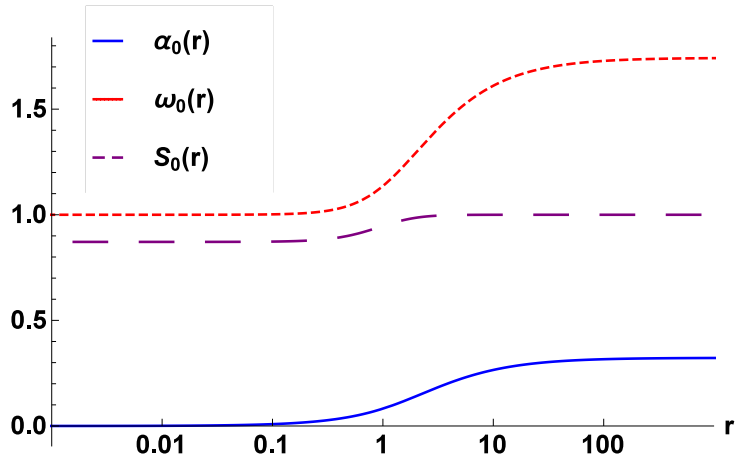}
\end{center}
\caption{Typical dyonic soliton solution with $\ell =1$, $\alpha _{1}=0.08715$
and $\omega _{2}=0.2$. We plot the electric gauge field function $\alpha _{0}(r)$ (blue, solid), magnetic gauge field function $\omega _{0}(r)$ (red, dotted) and metric function $S_{0}(r)$ (purple, dashed). Both the electric gauge field function
$\alpha _{0}(r)$ and the
magnetic gauge field function $\omega _{0}(r)$ are monotonically increasing and neither has any zeros for $r>0$.
The metric function $m_{0}(r)$ is not shown - it too is monotonically increasing.
}
\label{fig:one}
\end{figure}

\begin{figure}
\begin{center}
\includegraphics[width=10cm]{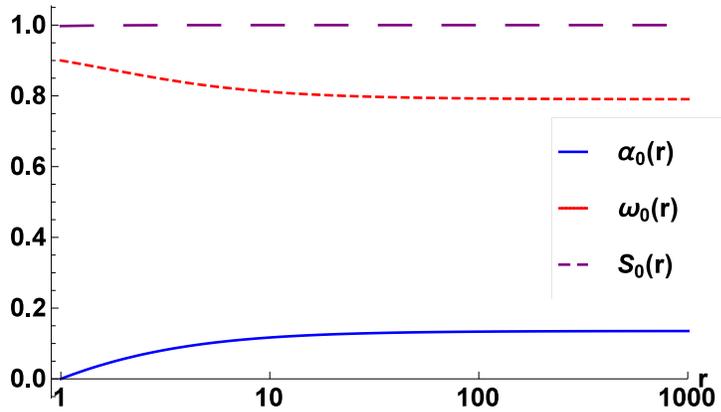}
\end{center}
\caption{Typical dyonic black hole solution with $\ell =1$, $r_{h}=1$,
$\alpha _{h}'=0.09974$ and $\omega _{h}=0.9$.  We plot the electric gauge field function $\alpha _{0}(r)$ (blue, solid), magnetic gauge field function $\omega _{0}(r)$ (red, dotted) and metric function $S_{0}(r)$ (purple, dashed).  The electric gauge field function $\alpha _{0}(r)$ is monotonically increasing and the magnetic gauge field  function $\omega _{0}(r)$ is monotonically decreasing. Neither gauge function has any zeros for $r>r_{h}$.
The function $S_{0}(r)$ varies only a little: its value on the horizon is $0.9974$, while at infinity $S_{0}(r) \rightarrow 1$.
The metric function $m_{0}(r)$ is not shown - it is monotonically increasing as $r$ increases.}
\label{fig:two}
\end{figure}

In \cite{Nolan} we have proven, for any value of the adS radius of curvature $\ell $, the existence of dyonic soliton and black hole solutions of the field equations (\ref{eq:static1}--\ref{eq:static4}) in a neighbourhood of the trivial (Schwarzschild-)adS space-time (\ref{eq:SadS}).
Providing the non-trivial solution is sufficiently close to the trivial solution, the magnetic gauge field function $\omega _{0}(r)$ will have no zeros.
In figures~\ref{fig:one} and \ref{fig:two} we show two typical nodeless solutions: a soliton and a black hole solution respectively.

More detailed properties of the space of dyonic solutions of ${\mathfrak {su}}(2)$ EYM in adS can be found in \cite{Bjoraker,Shepherd1}.
Our focus in this paper is the nodeless dyonic solitons and black holes.
In the case of purely magnetic solutions, it has been proven \cite{Bjoraker,Winstanley1} that at least some nodeless solitons and black holes are stable under linear, spherically symmetric, perturbations of the metric and gauge field.
We therefore expect that at least some nodeless dyonic solutions will also be stable under linear, spherically symmetric, perturbations.
In the next section we derive the equations governing such perturbations before proving the existence of stable dyonic solutions in section \ref{sec:stable}.

\section{Perturbation equations}
\label{sec:perteqns}

We now derive the equations satisfied by linear, time-dependent, spherically symmetric perturbations of the static equilibrium dyonic solitons and black holes
discussed in the previous section.

\subsection{Linearized perturbation equations}
\label{sec:linear}

We begin with the time-dependent field equations (\ref{eq:Einstein}--\ref{eq:YM}) for the metric (\ref{eq:metric}) and gauge potential (\ref{eq:gauge}).
The appropriateness of the form (\ref{eq:metric}) for studying perturbations of the static equilibrium solutions is discussed in section \ref{sec:gauge}.
The Einstein equations (\ref{eq:Einstein}) are
\numparts
\begin{eqnarray}
m' & = &
\frac {r^{2}}{2S^{2}} \left( {\dot {\beta }} - \alpha ' \right) ^{2} + \frac {1}{\mu S^{2}}\left[ {\dot {\omega }}^{2} +
\omega ^{2} \left( {\dot {\gamma }} + \alpha \right) ^{2} \right]
+ \mu \left[ \omega '^{2} + \omega ^{2} \left( \gamma ' +\beta \right) ^{2} \right]
\nonumber \\ & &
+ \frac {1}{2r^{2}} \left(  1 -\omega ^{2} \right) ^{2} ,
\label{eq:Einsteintime1}
\\
{\dot {m}} & = &
2\mu \left[ {\dot {\omega }} \omega ' + \omega ^{2} \left( {\dot {\gamma }} + \alpha \right) \left( \gamma '  + \beta \right) \right] ,
\\
\frac {S'}{S} & = &
\frac {2}{r} \left[ \omega '^{2} + \omega ^{2} \left( \gamma ' + \beta \right) ^{2} \right]
+\frac {2}{r\mu ^{2} S^{2}}\left[ {\dot {\omega }}^{2} + \omega ^{2} \left( {\dot {\gamma }} + \alpha \right) ^{2} \right] ,
\label{eq:Einsteintime2}
\end{eqnarray}
and the Yang-Mills equations (\ref{eq:YM}) take the form
\begin{eqnarray}
0 & = &
r^2 \mu \left( \alpha '' -{\dot {\beta }}' \right)
+\left( 2 r \mu -\frac{r^2 \mu S'}{S}\right) \left( \alpha ' -{\dot {\beta }} \right)
-2 \omega ^{2} \left( \alpha +  {\dot {\gamma }} \right) ,
\label{eq:YMtime1}
\\
0 & = &
r^{2}\left( {\ddot {\beta }} - {\dot {\alpha }}' \right)
- \frac {r^{2}{\dot {S}}}{S} \left( {\dot {\beta }} - \alpha ' \right)
+2\mu S^{2} \omega ^{2}\left( \beta +\gamma '\right) ,
\\
 0 & = &
 -\omega \left( {\ddot {\gamma }} + {\dot {\alpha }} \right)
 -2{\dot {\omega }} \left( {\dot {\gamma }}+ \alpha \right)
 + \frac {\left( \mu S  \right) {\dot {}}}{\mu S} \omega \left( {\dot {\gamma }}+ \alpha \right)
 + \left( \mu S \right) ^{2} \omega \left( \gamma '' + \beta ' \right)
 \nonumber \\ & &
 + \left[ 2 \left( \mu S \right) ^{2} \omega' + \mu S \left( \mu  S \right) ' \omega \right]  \left( \gamma ' + \beta \right) ,
\\
0 & = &
-{\ddot {\omega }} +  \frac {\left( \mu S \right) {\dot {}}}{\mu S }  {\dot {\omega }}
+ \omega \left( \alpha + {\dot {\gamma }} \right) ^{2}
+ \left( \mu  S \right) ^{2} \omega '' + \mu S \left( \mu S \right) ' \omega '
\nonumber \\ & &
- \mu ^{2} S^{2} \omega \left( \beta + \gamma ' \right) ^{2} + \frac {\mu S^{2}}{r^{2}} \left( 1- \omega ^{2}  \right) \omega ,
\label{eq:YMtime2}
\end{eqnarray}
where a dot ${\dot {}}$ denotes partial differentiation with respect to time $t$ and a prime $'$ denotes partial differentiation with respect to the
radial co-ordinate $r$.
\endnumparts

The ansatz for the gauge potential $A_{\tau }$ (\ref{eq:gauge}) possesses a residual ${\mathfrak {su}}(2)$ Lie algebra gauge freedom. If ${\mathfrak {g}}(t,r)$ is a diagonal $2\times 2$ matrix depending on $t$ and $r$, then the following gauge transformation leaves the form of the YM gauge potential (\ref{eq:gauge}) invariant, but changes the matrices
${\mathcal {A}}$, ${\mathcal {B}}$ and $C$:
\begin{eqnarray}
{\mathcal {A}} \rightarrow {\mathcal {A}}+{\mathfrak {g}}^{-1}
{\dot{{\mathfrak {g}}}} ,
\qquad
{\mathcal {B}}\rightarrow {\mathcal {B}}
+ {\mathfrak {g}}^{-1} {\mathfrak {g}}' ,
\nonumber \\
C-C^H\rightarrow {\mathfrak {g}}^{-1}\left( C-C^H \right) {\mathfrak {g}} ,
\qquad
C+C^H\rightarrow
{\mathfrak {g}}^{-1}\left( C+C^H \right) {\mathfrak {g}} .
\label{eq:gaugetrans}
\end{eqnarray}
Under this gauge transformation the YM gauge field strength $F_{\tau \nu }$ transforms as
\begin{equation}
F_{\tau \nu } \rightarrow {\mathfrak {g}}^{-1} F_{\tau \nu } {\mathfrak {g}}.
\end{equation}
In studies of the stability of purely magnetic ${\mathfrak {su}}(N)$ EYM solitons and black holes (see, for example, \cite{Baxter}),
the matrix ${\mathcal {A}}$ is identically equal to zero for the equilibrium solutions, and in this case the residual gauge freedom (\ref{eq:gaugetrans})
is used to set ${\mathcal {A}} \equiv 0$ for the time-dependent perturbations as well. Such a choice of gauge simplifies the analysis of the
resulting perturbation equations in that case.

Here we are interested in dyonic equilibrium solutions for which the matrix ${\mathcal {A}}$ does not vanish. In this case the most appropriate choice of Lie algebra gauge is not immediately apparent.  Instead of choosing a gauge,
we consider gauge-invariant variables which do not change under the Lie algebra gauge transformation (\ref{eq:gaugetrans}).
The gauge-invariant variables we use are:
\begin{equation}
\psi = \alpha ' - {\dot {\beta }}
\qquad
\xi = {\dot {\gamma }} + \alpha ,
\qquad
\eta = \gamma ' + \beta ,
\label{eq:gaugeinvs}
\end{equation}
together with the variable $\omega $ which is unchanged by the gauge transformation (\ref{eq:gaugetrans}).
We can also eliminate the gauge-invariant variable $\psi $ since
\begin{equation}
\psi = \xi ' - {\dot {\eta }}.
\end{equation}
With these new variables, the Einstein equations (\ref{eq:Einsteintime1}--\ref{eq:Einsteintime2}) take the more compact form
\numparts
\begin{eqnarray}
m' & = &
\frac {r^{2}}{2S^{2}} \left( \xi ' - {\dot {\eta }} \right) ^{2} + \frac {1}{\mu S^{2}} \left( {\dot {\omega }}^{2} + \omega ^{2} \xi ^{2} \right)
+ \mu \left( \omega '^{2} + \omega ^{2} \eta ^{2} \right)
\nonumber \\ & &
+ \frac {1}{2r^{2}} \left( 1- \omega ^{2}\right) ^{2} ,
\label{eq:Einsteingauge1}
\\
{\dot {m}} & = &
2\mu \left( {\dot {\omega }}\omega ' + \omega ^{2} \xi \eta \right) ,
\\
\frac {S'}{S} & = &
\frac {2}{r} \left( \omega '^{2} + \omega ^{2} \eta ^{2} \right)
+ \frac {2}{r\mu ^{2} S^{2}} \left( {\dot {\omega }}^{2} + \omega ^{2} \xi ^{2} \right) ,
\label{eq:Einsteingauge2}
\end{eqnarray}
and the Yang-Mills equations (\ref{eq:YMtime1}--\ref{eq:YMtime2}) also simplify:
\begin{eqnarray}
0 & = &
\mu  \left( \xi '' - {\dot {\eta }}'  \right) + \left( \frac {2\mu }{r} - \frac {\mu S'}{S} \right) \left( \xi ' -{\dot {\eta }} \right) - \frac {2\omega ^{2}}{r^{2}} \xi ,
\label{eq:YMgauge1}
\\
0 & = &
-{\ddot {\eta }} + {\dot {\xi }}' - \frac {{\dot {S}}}{S} \left( \xi ' - {\dot {\eta }} \right) - \frac {2\mu S^{2} \omega ^{2}}{r^{2}} \eta ,
\\
0 & = &
-\omega {\dot {\xi }} - 2{\dot {\omega }} \xi  + \left( \mu S\right) ^{2} \left( \omega \eta ' + 2\omega ' \eta \right)
+ \frac {\left( \mu S \right) {\dot {}}}{\mu  S} \omega \xi
+ \mu S \left(  \mu  S\right) ' \omega \eta ,
\\
0 & = &
-{\ddot {\omega }} + \frac {\left( \mu S \right) {\dot {}}}{\mu S} {\dot {\omega }}
+ \omega  \xi ^{2} + \left( \mu S\right) ^{2} \omega ''
+ \mu S \left( \mu S  \right) ' \omega '
-\left( \mu S\right) ^{2} \omega \eta ^{2}
\nonumber \\ & &
 + \frac {\mu S^{2} \omega }{r^{2}} \left(  1- \omega ^{2} \right) .
\label{eq:YMgauge2}
\end{eqnarray}
\endnumparts

We now consider linearized perturbations about the static equilibrium solutions discussed in section~\ref{sec:static}.
The field variables are written as sums of the time-independent equilibrium quantities (denoted by a subscript $0$, for example $\mu _{0}(r)$)
and small time-dependent perturbations (denoted by a $\delta $, for example $\delta \mu (t,r)$).
For the static equilibrium solutions, we have $\beta _{0}= \gamma _{0}=0$ and hence, for the gauge-invariant variables (\ref{eq:gaugeinvs}), we have
\begin{equation}
\xi _{0}(r) = \alpha _{0}(r), \qquad
\eta _{0}(r) = 0.
\end{equation}
The time-dependent field variables are therefore written as follows
\begin{eqnarray}
\mu (t,r) & = & \mu _{0} (r) + \delta \mu (t,r) , \qquad
S(t,r)  =  S_{0}(r) + \delta S (t,r),
\nonumber \\
\xi (t,r) & = &  \alpha _{0}(r) + \delta \xi (t,r), \qquad
\eta (t,r)   =   \delta \eta (t,r) ,
\nonumber \\
\omega (t,r) &  = &  \omega _{0}(r) + \delta \omega (t,r) .
\label{eq:perts}
\end{eqnarray}
The field variables (\ref{eq:perts}) are substituted into the field equations (\ref{eq:Einsteingauge1}--\ref{eq:YMgauge2}), working to first order
in the perturbations.
The resulting equations are simplified using the equilibrium field equations (\ref{eq:static1}--\ref{eq:static4}).
The linearized perturbed Einstein equations are then
\numparts
\begin{eqnarray}
\delta \mu ' & = &
 \frac {1}{r} \left( \frac {2\alpha _{0}^{2}\omega _{0}^{2}}{\mu _{0}^{2} S_{0}^{2}} - 2\omega _{0}'^{2}  -1 \right) \delta \mu
+ \frac {2}{S_{0}^{3}} \left( \frac {2\alpha _{0}^{2}\omega _{0}^{2}}{r \mu _{0}} + r\alpha _{0}'^{2} \right) \delta S
\nonumber \\ & &
- \frac {4\mu _{0} \omega _{0}'}{r} \delta \omega '
+ \frac {4\omega _{0}}{r} \left( \frac {1 - \omega _{0}^{2}}{r^{2}} - \frac {\alpha _{0}^{2}}{\mu _{0}S_{0}^{2}}  \right) \delta \omega
- \frac {2r\alpha _{0}'}{S_{0}^{2}} \left( \delta \xi ' - \delta {\dot {\eta }} \right)
\nonumber \\ & &
- \frac {4\alpha _{0}\omega _{0}^{2}}{r\mu _{0} S_{0}^{2}} \delta \xi ,
\label{eq:EEpert1}
\\
\delta {\dot {\mu }} & = &
-\frac {4\mu _{0} }{r} \left(
 \omega _{0}' \delta {\dot {\omega }}
+ \alpha _{0} \omega _{0} ^{2} \delta \eta \right) ,
\label{eq:EEpert2}
\\
\delta S ' & = &
- \frac {4 \alpha _{0}^{2}\omega _{0}^{2}}{r\mu _{0}^{3} S_{0}} \delta \mu
+ \frac {2}{r} \left( \omega _{0}'^{2} - \frac {\alpha _{0}^{2}\omega _{0}^{2}}{\mu _{0}^{2} S_{0}^{2}}  \right) \delta S
+ \frac {4S_{0}\omega _{0}'}{r} \delta \omega '
+ \frac {4\alpha _{0}^{2} \omega _{0}}{r\mu _{0}^{2} S_{0}} \delta \omega
\nonumber \\ & &
+\frac {4\alpha _{0}\omega _{0}^{2}}{r\mu _{0}^{2} S_{0}} \delta \xi,
\label{eq:EEpert3}
\end{eqnarray}
and the perturbed Yang-Mills equations take the form
\begin{eqnarray}
0 & = &
\mu _{0} \left( \delta \xi '' - \delta {\dot {\eta }}' \right)
+ \frac {2\alpha _{0}\omega _{0}^{2}}{r^{2}\mu _{0}} \delta \mu
-  \frac {\mu _{0}\alpha _{0}'}{S_{0}} \delta S'
+ \frac {\mu _{0}\alpha _{0}' S_{0}'}{S_{0}^{2}} \delta S
- \frac {4\alpha _{0}\omega _{0}}{r^{2}} \delta \omega
\nonumber \\ & &
+ \mu _{0} \left( \frac {2}{r} - \frac {S_{0}'}{S_{0}} \right) \left( \delta \xi' - \delta {\dot {\eta }} \right)
- \frac {2\omega _{0}^{2}}{r^{2}} \delta \xi ,
\label{eq:YMpert1}
\\
0 & = &
-\delta {\ddot {\eta }} + \delta {\dot {\xi }}'
- \frac {\alpha _{0}' }{S_{0}} \delta {\dot {S}}
-\frac {2\mu _{0} S_{0}^{2}\omega _{0}^{2}}{r^{2}} \delta \eta ,
\label{eq:YMpert2}
\\
0 & = &
\omega _{0} \delta {\dot {\xi }}
+ 2\alpha _{0}  \delta {\dot {\omega }}
-\alpha _{0}\omega _{0} \left(
\frac{ \delta {\dot {\mu }} }{\mu _{0}}
+  \frac {\delta {\dot {S}}}{S_{0}}  \right)
- \mu _{0}^{2}S_{0}^{2} \omega _{0} \delta \eta '
\nonumber \\ & &
-\mu _{0} S_{0} \left[
2\mu _{0} S_{0} \omega _{0}'
+ \left( \mu _{0} S_{0} \right)' \omega _{0}
\right] \delta \eta  ,
\label{eq:YMpert3}
\\
0 & = &
- \delta {\ddot {\omega }}
+ \mu _{0}^{2} S_{0}^{2} \delta \omega ''
+ \mu _{0}S_{0} \left( \mu _{0} S_{0}\right) ' \delta \omega '
+ \mu _{0} S_{0} \omega _{0}' \left( S_{0}\delta \mu '+ \mu _{0} \delta S' \right)
\nonumber \\ & &
- S_{0}^{2} \left( \mu _{0}' \omega _{0}'
+ \frac {\omega _{0}\left( 1- \omega _{0}^{2}\right)}{r^{2}} + \frac {2\alpha _{0}^{2}\omega _{0}}{\mu _{0}S_{0}^{2}} \right) \delta \mu
- \left( \frac {2\alpha _{0}^{2}\omega _{0}}{S_{0}}+ \mu _{0}^{2} S_{0}' \omega _{0}' \right) \delta S
\nonumber \\ & &
+ \left( \alpha _{0}^{2} + \frac {\mu _{0} S_{0}^{2} \left( 1 - 3\omega _{0}^{2} \right) }{r^{2}} \right) \delta \omega
+ 2\alpha _{0}\omega _{0}\delta \xi .
\label{eq:YMpert4}
\end{eqnarray}
\endnumparts

\subsection{Space-time diffeomorphism gauge transformations}
\label{sec:gauge}

In addition to gauge invariance with respect to representations of the ${\mathfrak{su}(2)}$ Lie algebra as discussed above, we must also consider the question of gauge invariance with respect to space-time diffeomorphisms. This is relevant when dealing with perturbations of a space-time, as we must ensure that the quantities encountered are indeed genuine perturbations, and not just artefacts of an infinitesimal co-ordinate transformation carried out on the background space-time. The key idea is to consider the effect of such an infinitesimal co-ordinate transformation generated by the vector field $V^{\tau }$, such that
\begin{equation}
x^\tau \rightarrow x^\tau+ V^\tau.
\label{eq:coord-transf}
\end{equation}
Under such a transformation, the metric perturbation $\delta g_{\tau \nu}$ changes according to
\begin{equation}
\delta g_{\tau \nu} \rightarrow \delta g_{\tau \nu} + {\mathcal{L}}_{\vec{V}}g_{\tau \nu},
\label{eq:lie-metric}
\end{equation}
where ${\mathcal{L}}_{\vec{V}}$ is the Lie derivative along $\vec{V}$. Likewise, the gauge potential perturbation (an $\mathfrak{su}(2)$-valued one-form) undergoes the transformation
\begin{equation}
\delta A_\tau \rightarrow \delta A_\tau + {\cal{L}}_{\vec{V}}A_\tau .
\label{eq:lie-potential}
\end{equation}

We are considering time-dependent, spherically symmetric perturbations of the static, equilibrium configurations. Unlike the situation that holds for non-spherical perturbations (decomposed into multipoles of appropriate valence), we do not have a complete set of gauge-invariant quantities to work with \cite{Gerlach-Sengupta}. We must therefore be cautious in identifying genuine perturbations, and perturbations which are pure gauge. We must also ensure that we are considering perturbations of maximal generality. In the co-ordinates $(t,r)$, the most general bare spherically symmetric perturbation of the metric has the form
\begin{equation}
\delta g_{\tau \nu} = \left( \begin{array}{cc} \begin{array}{cc} \delta g_{00} & \delta g_{01} \\ \delta g_{01} & \delta g_{11} \end{array} & 0_2 \\
0_2 & \begin{array}{cc} \delta g_{22} & 0 \\ 0 & \delta g_{22}\sin^2\theta \end{array} \end{array} \right),
\label{eq:del-metric}
\end{equation}
where $0_2$ is the $2\times 2$ zero matrix. Under the gauge transformation (\ref{eq:lie-metric}) generated by
\begin{equation}
V^\tau = (x(t,r), y(t,r),0,0),
\label{eq:ss-gauge-vector}
\end{equation}
we have
\numparts
\begin{eqnarray}
\delta g_{01} &\rightarrow  & \delta g_{01} - \mu_0S_0^2x' + \mu_0^{-1}\dot{y},
\label{eq:gt-g01}\\
\delta g_{22} &\rightarrow & \delta g_{22} + 2ry.
\label{eq:gt-g22}
\end{eqnarray}
\endnumparts

We exploit these to simplify the form of the perturbation as follows. We begin with the completely general perturbation (\ref{eq:del-metric}). We then make a diffeomorphism gauge transformation generated by (\ref{eq:ss-gauge-vector}), choosing
\begin{equation}
y = -\frac{\delta g_{22}}{2r}.
\end{equation}
This brings us to a gauge in which
\begin{equation}
\delta g_{22}=0.
\end{equation}
This condition is preserved by further gauge transformations provided that $y=0$ in (\ref{eq:ss-gauge-vector}). We then apply such a further transformation, choosing $\vec{V}$ so that
\begin{equation}
x' = \frac{\delta g_{01}}{\mu_0S_0^2}.
\end{equation}
This yields a gauge in which
\begin{equation}
\delta g_{01} = \delta g_{22} = 0,
\label{eq:diag-gauge}
\end{equation}
which is preserved by {\textit{further}} gauge transformations generated by gauge vectors of the form
\begin{equation}
V^\tau = \left(x(t),0,0,0 \right).
\label{eq:gauge-freedom}
\end{equation}
This represents the generator of the only gauge freedom that remains in the problem, and corresponds to a redefinition of the time co-ordinate via (\ref{eq:coord-transf}).
We will refer to the gauge condition (\ref{eq:diag-gauge}) as the {\textit{diagonal gauge}}, and we note that the perturbed metric now has the form
\begin{equation}
\delta g_{\tau \nu} = \Diag(\delta g_{00}, \delta g_{11}, 0, 0).
\label{eq:del-metric-diag}
\end{equation}
Thus the metric perturbation may be represented by a perturbation of the background metric functions:
\begin{equation}
\mu_0(r) \to \mu_0(r) + \delta\mu(t,r),\qquad S_0(r) \to S_0(r) + \delta S(t,r),
\label{eq:del-metric-fns}
\end{equation}
as assumed in the previous subsection.

For perturbations of static, spherically symmetric space-times, the choice of diagonal gauge (and the reasons behind this choice) are standard and well-known.
However, for our analysis it is important to understand the residual diffeomorphism gauge freedom given by (\ref{eq:gauge-freedom}), in particular its effect on the metric and matter perturbations.
These are summarized in the following lemma.
\begin{Lemma}
The most general spherically symmetric perturbation of the metric (\ref{eq:metric}) may be written in the diagonal gauge (\ref{eq:del-metric-diag}) where
\begin{equation}
\delta g_{00} = -S_0^2 \delta\mu -2\mu_0S_0\delta S, \qquad
\delta g_{11} = -\mu_0^{-2} \delta\mu.
\end{equation}
Under the remaining gauge freedom of infinitesimal co-ordinate transformations generated by (\ref{eq:gauge-freedom}), the metric perturbation functions transform as
\numparts
\begin{equation}
\delta \mu \rightarrow \delta \mu, \qquad
\delta S \rightarrow \delta S +S_0\dot{x} .
\label{eq:gf-del-s}
\end{equation}
Furthermore, we can deduce from (\ref{eq:lie-potential}) that the matter perturbations transform as
\begin{equation}
\delta \omega \rightarrow \delta\omega, \quad
\delta \psi \rightarrow \delta \psi + \alpha_0'\dot{x}, \quad
\delta \xi \rightarrow \delta \xi + \alpha_0\dot{x}, \quad
\delta \eta \rightarrow \delta \eta.
\label{eq:gf-del-xi}
\end{equation}
\endnumparts
\end{Lemma}
The results of this lemma, in particular the behaviour of $\delta S $ (\ref{eq:gf-del-s}) and $\delta \xi $ (\ref{eq:gf-del-xi}),  will be useful in the next subsection for understanding the final form of the linearized perturbation equations.

\subsection{Linearized perturbation equations in standard form}
\label{eq:perttidy}

We now seek to set the linearized perturbation equations (\ref{eq:EEpert1}--\ref{eq:YMpert4}) into a form amenable to proving the existence of stable
equilibrium solutions.
From (\ref{eq:EEpert2}, \ref{eq:YMpert2}, \ref{eq:YMpert3}) it can be seen that the perturbation $\delta \eta $ is out of phase with the other perturbations: when this quantity appears in the perturbation equations with an even (respectively, odd) number of time derivatives, all other variables appear with an odd (respectively, even) number of time derivatives. (In physical terms, this means that if $\delta\eta$ follows a sine wave, the other variables follow a cosine wave, and vice versa.) We therefore define a new quantity $\delta \kappa $ by
\begin{equation}
\delta \eta (t,r) = \frac {\delta {\dot {\kappa }}(t,r)}{\mu _{0}S_{0}\omega _{0}}.
\label{eq:kappadef}
\end{equation}
We note that $\delta \kappa $ is defined only up to an arbitrary function of the radial co-ordinate $r$, and this freedom in defining $\delta \kappa $ will be useful in our later analysis.
The equilibrium functions of $r$ are introduced in (\ref{eq:kappadef}) because they will enable us to ultimately set the perturbation equations into a standard form.
With the substitution (\ref{eq:kappadef}), equations (\ref{eq:EEpert2}, \ref{eq:YMpert2}, \ref{eq:YMpert3}) then take the form
\numparts
\begin{eqnarray}
\delta {\dot {\mu }} & = &
-\frac {4}{rS_{0}} \left(
\mu_{0} S_{0} \omega _{0}' \delta {\dot {\omega }}
+ \alpha _{0} \omega _{0} \delta {\dot {\kappa }} \right) ,
\label{eq:EEpert2kappa}
\\
0 & = &
\frac {\delta {\dddot {\kappa}}}{\mu _{0}S_{0}\omega _{0}}
+\frac {2S_{0}\omega _{0}}{r^{2}} \delta {\dot {\kappa }}
+ \frac {\alpha _{0}' }{S_{0}} \delta {\dot {S}}
- \delta {\dot {\xi }}' ,
\label{eq:YMpert2kappa}
\\
0 & = &
\delta {\dot {\kappa }}' + \frac {\omega _{0}'}{\omega _{0}} \delta {\dot {\kappa }}
+ \frac {\alpha _{0}\omega _{0}}{\mu _{0} S_{0}} \left( \frac {\delta {\dot {\mu }}}{\mu _{0}} + \frac {\delta {\dot {S}}}{S_{0}} \right)
-\frac {2\alpha _{0}}{\mu _{0}S_{0}} \delta {\dot {\omega }}
- \frac {\omega _{0}}{\mu _{0}S_{0}} \delta {\dot {\xi }}.
\label{eq:YMpert3kappa}
\end{eqnarray}
\endnumparts
Integrating (\ref{eq:EEpert2kappa}) with respect to time gives the metric perturbation $\delta \mu (t,r)$ to be
\begin{equation}
\delta \mu   =
-\frac {4}{rS_{0}} \left(
\mu_{0} S_{0} \omega _{0}' \delta \omega
+ \alpha _{0} \omega _{0} \delta \kappa \right)
+ \delta {\mathcal {F}}(r),
\label{eq:deltamu}
\end{equation}
where $\delta {\mathcal {F}}(r)$ is an arbitrary function of $r$.
For purely magnetic background solutions with $\alpha _{0}\equiv 0$, the expression (\ref{eq:deltamu}) reduces to that in \cite{Baxter} for this metric perturbation.

Substituting for $\delta \mu (t,r)$ (\ref{eq:deltamu}) in (\ref{eq:YMpert3kappa}) and integrating with respect to time gives the other metric perturbation $\delta S(t,r)$:
\begin{eqnarray}
\delta S & = &
-\frac {\mu _{0}S_{0}^{2}}{\alpha _{0}\omega _{0}}\delta \kappa '
+ \left(
\frac {4\alpha _{0}\omega _{0}}{r\mu _{0}} - \frac {\mu _{0}S_{0}^{2} \omega _{0}'}{\alpha _{0}\omega _{0}^{2}}
 \right) \delta \kappa
+ 2S_{0} \left( \frac {1}{\omega _{0}} + \frac {2\omega _{0}'}{r}  \right) \delta \omega
\nonumber \\ & &
+\frac {S_{0}}{\alpha _{0}} \delta \xi
+\delta {\mathcal {G}}(r),
\label{eq:deltaS}
\end{eqnarray}
where $\delta {\mathcal {G}}(r)$ is another arbitrary function of $r$.
We note that this expression is not valid for purely magnetic background solutions with $\alpha _{0} \equiv 0$.
When the background solutions are purely magnetic, the perturbed Einstein equations can be used to find an expression for $\delta S'$ but not $\delta S $
\cite{Baxter}.

Substituting for $\delta S (t,r)$ in (\ref{eq:YMpert2kappa}) and integrating with respect to time gives the following:
\begin{eqnarray}
0 & = &
\delta {\ddot {\kappa }}
- \frac {\left( \mu _{0} S_{0} \right) ^{2} \alpha _{0}'}{\alpha _{0}} \delta \kappa '
- \mu _{0} S_{0} \omega _{0} \delta \xi '
+ 2\mu _{0} S_{0} \alpha _{0}' \left(
1 + \frac {2\omega _{0}\omega _{0}'}{r}
\right) \delta \omega
\nonumber \\ & &
+ \frac {\mu _{0}S_{0} \alpha _{0}'\omega _{0}}{\alpha _{0}} \delta \xi
+ \left(
\frac {2\mu _{0}S_{0}^{2}\omega _{0}^{2}}{r^{2}}
+ \frac {4\alpha _{0} \alpha _{0}'\omega _{0}^{2}}{r}
- \frac {\mu _{0}^{2} S_{0}^{2} \alpha _{0}' \omega _{0}'}{\alpha _{0}\omega _{0}}
\right) \delta \kappa
+ \delta {\mathcal {H}}(r),
\nonumber \\
\label{eq:kappaddot1}
\end{eqnarray}
where $\delta {\mathcal {H}}(r)$ is a third arbitrary function of $r$.
A second equation involving $\delta {\ddot {\kappa }}$ can be derived from (\ref{eq:EEpert1}), substituting in for $\delta {\dot {\eta}}$ using (\ref{eq:kappadef}), for $\delta \mu $ using (\ref{eq:deltamu}) and for $\delta S$ using (\ref{eq:deltaS}).
Subtracting the resulting equation from (\ref{eq:kappaddot1}) gives a constraint on the arbitrary functions
$\delta {\mathcal {F}}(r)$, $\delta {\mathcal {G}}(r)$ and $\delta {\mathcal {H}}(r)$:
\begin{eqnarray}
0 & = & \delta {\mathcal {F}}'
- \left(
\frac {2\alpha _{0}^{2}\omega _{0}^{2}}{\mu _{0}^{2}S_{0}^{2}} -1 - 2\omega _{0}'^{2}
\right) \frac { \delta {\mathcal {F}} }{r}
- \frac {2r\alpha _{0}'}{\mu _{0}S_{0}^{3}} \left(
\mu _{0} \alpha _{0}' + \frac {2\alpha _{0}^{2}\omega _{0}^{2}}{r^{2}\alpha _{0}'}
\right)\delta {\mathcal {G}}
\nonumber \\ & &
+\frac {2r\alpha _{0}'}{\mu _{0}S_{0}^{3}\omega _{0}}\delta {\mathcal {H}}.
\label{eq:constraint1}
\end{eqnarray}

A second, independent, constraint on the functions $\delta {\mathcal {F}}(r)$, $\delta {\mathcal {G}}(r)$ and $\delta {\mathcal {H}}(r)$ is derived from
(\ref{eq:YMpert1}) as follows. First rearrange the equation resulting from integrating (\ref{eq:YMpert2kappa}) with respect to time to give an expression for $\delta S$, which involves both $\delta {\mathcal {G}}$ and $\delta {\mathcal {H}}$.
Next differentiate this with respect to $r$ and then substitute into (\ref{eq:YMpert1}), simplifying using the forms (\ref{eq:deltamu}, \ref{eq:deltaS}) of the metric perturbations.
Using (\ref{eq:kappaddot1}) to eliminate $\delta {\ddot {\kappa }}$ from the resulting equation gives the constraint, whose most compact form reads
\begin{eqnarray}
0 & = &
\frac {2 S_{0} \alpha _{0} \omega _{0}^{2}}{r^{2} \mu _{0}^{2} \alpha _{0}'} \delta {\mathcal {F}}
- \delta {\mathcal {G}}'
+ \frac {S_{0}'}{S_{0}} \delta {\mathcal {G}}
+ \frac {1}{\mu _{0} \alpha _{0}' \omega _{0}} \delta {\mathcal {H}}'
\nonumber \\ & &
+ \frac {1}{\mu _{0} \alpha _{0}' \omega _{0}} \left(
\frac {2}{r} - \frac {\mu _{0}'}{\mu _{0}} -\frac {2S_{0}'}{S_{0}} - \frac {\omega _{0}'}{\omega _{0}}
 \right) \delta {\mathcal {H}} .
\label{eq:constraint2}
\end{eqnarray}

Next we use the remaining perturbed Einstein equation (\ref{eq:EEpert3}) to give a first order equation for $\delta \xi $, which involves only derivatives with respect to $r$:
\begin{eqnarray}
0 & = &
\delta \xi ' - \frac {\alpha _{0}'}{\alpha _{0}} \delta \xi
-\frac {\mu _{0}S_{0}}{\omega _{0}} \delta \kappa ''
+ \frac {2\alpha _{0}}{\omega _{0}} \delta \omega '
+ \frac {\mu _{0}S_{0}}{\omega _{0}} \left(
\frac {\alpha _{0}'}{\alpha _{0}} - \frac {\mu _{0}'}{\mu _{0}}  - \frac {S_{0}'}{S_{0}}
\right) \delta \kappa '
\nonumber \\ & &
-\frac{ 4 \alpha _{0} \omega _{0}' }{r} \left(
\frac {\omega _{0}\left( 1 - \omega _{0}^{2} \right) }{r^{2}\mu _{0}\omega _{0}'}
+ \frac {1}{r}
+ \frac {r}{2\omega _{0}^{2}}
+ \frac {\mu _{0}'}{\mu _{0}}
+ \frac {S_{0}'}{S_{0}}
 \right) \delta \omega
\nonumber \\ & &
+ \frac {4\alpha _{0}^{2}\omega _{0}}{r\mu _{0}S_{0}} \left(
\frac {r}{4\omega _{0}^{2}}
+ \frac {\mu_{0}S_{0}^{2}\left( 1- \omega _{0}^{2} \right)}{4r\alpha _{0}^{2} \omega _{0}^{2}}
- \frac {1}{r}
+ \frac {\alpha _{0}'}{\alpha _{0}}
- \frac {\mu _{0}'}{\mu _{0}} -\frac {S_{0}'}{S_{0}}
\right. \nonumber \\ & & \left.
+ \frac {r\mu _{0}^{2}S_{0}^{2} \alpha _{0}'\omega _{0}'}{4\alpha _{0}^{3}\omega _{0}^{3}}
+ \frac {r\mu _{0}^{2}S_{0}^{2} \omega _{0}'^{2}}{2\alpha _{0}^{2}\omega _{0}^{4}}
 \right) \delta \kappa
+\frac {4\alpha _{0}^{3} \omega _{0}^{2}}{r\mu _{0}^{3}S_{0}^{2}} \delta {\mathcal {F}}
+\frac {\alpha _{0}}{S_{0}} \delta {\mathcal {G}}'
\nonumber \\ & &
+ \frac {2\alpha _{0}}{rS_{0}} \left(
\frac {\alpha _{0}^{2} \omega _{0}^{2}}{\mu _{0}^{2} S_{0}^{2}} -\omega _{0}'^{2}
\right) \delta {\mathcal {G}} .
\label{eq:deltaxi}
\end{eqnarray}
We use (\ref{eq:deltaxi}) to eliminate $\delta \xi $ from (\ref{eq:kappaddot1}) and the remaining perturbed Yang-Mills equation (\ref{eq:YMpert4}).
In terms of the usual ``tortoise'' co-ordinate $r_{*}$, defined by
\begin{equation}
\frac {dr_{*}}{dr} = \frac {1}{\mu _{0}S_{0}},
\label{eq:tortoise}
\end{equation}
the resulting pair of coupled perturbation equations takes the form
\numparts
\begin{eqnarray}
-\delta {\ddot {\kappa }} & = &
-\partial _{r_{*}}^{2} \delta \kappa
+ 2\alpha _{0} \partial _{r_{*}} \delta \omega
+ {\mathcal {E}}_{1} \delta \kappa + {\mathcal {E}}_{2} \delta \omega
+ \frac {4\alpha _{0}^{3} \omega _{0}^{3}}{r\mu _{0}^{2} S_{0}} \delta {\mathcal {F}}
\nonumber \\ & &
+ \mu _{0} \alpha _{0} \omega _{0} \delta {\mathcal {G}}'
+ \frac {2\mu _{0}\alpha _{0} \omega _{0}}{r} \left(
\frac {\alpha _{0}^{2} \omega _{0}^{2}}{\mu _{0}^{2} S_{0}^{2}} - \omega _{0}'^{2}
\right) \delta {\mathcal {G}}
+ \delta {\mathcal {H}},
\label{eq:pair1}
\\
-\delta {\ddot {\omega }} & = &
- \partial _{r_{*}}^{2} \delta \omega
- 2\alpha _{0} \partial _{r_{*}} \delta \kappa
+ {\mathcal {E}}_{3} \delta \kappa + {\mathcal {E}}_{4} \delta \omega
\nonumber \\ & &
-\mu _{0} S_{0}^{2} \omega _{0}' \delta {\mathcal {F}}'
+ \frac {2 \alpha _{0}^{2}\omega _{0}}{rS_{0}} \left(
r + 2 \omega _{0} \omega _{0}'
\right) \delta {\mathcal {G}}
\nonumber \\ & &
+ \left[
\frac {S_{0}^{2}\omega _{0} \left( 1 - \omega _{0}^{2} \right) }{r^{2}}
+ \frac {2\alpha _{0}^{2} \omega _{0}}{r\mu _{0}}
\left( r + 2\omega _{0} \omega _{0}' \right)
+ \mu _{0}'S_{0}^{2} \omega _{0}'
\right] \delta {\mathcal {F}},
\label{eq:pair2}
\end{eqnarray}
\endnumparts
where a prime $'$ denotes differentiation with respect to the radial co-ordinate $r$, and the coefficients ${\mathcal {E}}_{1}$, ${\mathcal {E}}_{2}$,
${\mathcal {E}}_{3}$ and ${\mathcal {E}}_{4}$ are given by
\numparts
\begin{eqnarray}
{\mathcal {E}}_{1}  & = &
\alpha _{0}^{2}
+ \frac {\mu _{0} S_{0}^{2} \left( 1 + \omega _{0}^{2} \right) }{r^{2}}
+ \frac {2\mu _{0}^{2}S_{0}^{2} \omega _{0}'^{2}}{\omega _{0}^{2}}
+ \frac {4\alpha _{0}^{2} \omega _{0}^{2}}{r}
\left(  \frac {2\alpha _{0}'}{\alpha _{0}}
-\frac {1}{r} - \frac {\mu _{0}'}{\mu _{0}} - \frac {S_{0}'}{S_{0}}
\right) ,
\nonumber \\ & &
\label{eq:E1}
\\
{\mathcal {E}}_{2} & = &
2 \mu _{0}S_{0} \alpha _{0}'
- \frac {4S_{0}\alpha _{0} \omega _{0}^{2} \left( 1 - \omega _{0}^{2} \right) }{r^{3}}
\nonumber \\ & &
+ \frac {4\mu _{0}S_{0} \alpha _{0} \omega _{0} \omega _{0}'}{r}
\left(
\frac {\alpha _{0}'}{\alpha _{0}}
-\frac {1}{r} -\frac {r}{2\omega _{0}^{2}}
- \frac {\mu _{0}'}{\mu _{0}}
- \frac {S_{0}'}{S_{0}}
\right) ,
\label{eq:E2}
\\
{\mathcal {E}}_{3} & = &
{\mathcal {E}}_{2} - 2\mu _{0} S_{0}\alpha _{0}' ,
\label{eq:E3}
\\
{\mathcal {E}}_{4} & = &
3\alpha _{0}^{2}
-\frac {\mu _{0}S_{0}^{2}\left( 1 - 3\omega _{0}^{2} \right) }{r^{2}}
- \frac {4\mu _{0}^{2}S_{0}^{2}\omega _{0}'^{2}}{r}
 \left(
 \frac {\mu _{0}'}{\mu _{0}}
 +\frac {S_{0}' }{S_{0}}
\right)
\nonumber \\  & &
- \frac {4\mu _{0}S_{0}^{2}\omega _{0}'}{r^{3}}
\left[
 2\omega _{0} \left( 1 - \omega _{0}^{2} \right)
+ r \mu _{0} \omega _{0}' \right] .
\label{eq:E4}
\end{eqnarray}
\endnumparts
For purely magnetic background solutions with $\alpha _{0}\equiv 0$, we have ${\mathcal {E}}_{2}\equiv 0 \equiv {\mathcal {E}}_{3}$
and the equations (\ref{eq:pair1}, \ref{eq:pair2}) are no longer coupled, instead giving separate equations for $\delta \kappa $ and $\delta \omega $.
In this case ${\mathcal {E}}_{1}$ reduces to the potential in \cite{Winstanley1} for sphaleronic sector perturbations of the purely magnetic background solutions, while ${\mathcal {E}}_{4}$ reduces to the potential in \cite{Winstanley1} for gravitational sector perturbations of the purely magnetic background solutions.
For black hole solutions, the boundary conditions (\ref{eq:horizon}) ensure that ${\mathcal {E}}_{i}$, $i=1,\ldots 4$ all vanish as  $r\rightarrow r_{h}$.
For soliton solutions, as $r\rightarrow 0$ the functions ${\mathcal {E}}_{2}$ and ${\mathcal {E}}_{3}$ tend to constants; however ${\mathcal {E}}_{1}$ and ${\mathcal {E}}_{4}$ diverge like a positive constant multiplied by $r^{-2}$.
At infinity, they all tend to constants.
The functions ${\mathcal {E}}_{i}$, $i=1,\ldots 4$ are all regular for $r>0$ or $r\ge r_{h}$, as applicable, provided that the equilibrium magnetic gauge field function $\omega _{0}(r)$ has no zeros. Our stability analysis in the next section will be applicable only to equilibrium solutions for which this is the case.

Returning to the case of dyonic background solutions with nontrivial $\alpha _{0}$, we have two coupled, dynamical, perturbation equations (\ref{eq:pair1}, \ref{eq:pair2}) for the perturbations $\delta \kappa $ and $\delta \omega $, together with a constraint equation (\ref{eq:deltaxi}) for $\delta \xi $, which does not contain any time derivatives.  We also have in our system three arbitrary functions of the radial co-ordinate $r$ only, namely ${\mathcal {F}}$,
${\mathcal {G}}$ and ${\mathcal {H}}$, which are constrained by the equations (\ref{eq:constraint1}, \ref{eq:constraint2}).
We now argue that all three of these can, without loss of generality, be set equal to zero.

We start by noting that the variable $\delta \kappa $ is defined by (\ref{eq:kappadef}) only up to an arbitrary function of $r$.
This freedom in the definition of $\delta \kappa $ can be used, via (\ref{eq:deltamu}), to set $\delta {\mathcal {F}}\equiv 0$.
The first constraint equation (\ref{eq:constraint1}) then gives $\delta {\mathcal {H}}$ in terms of $\delta {\mathcal {G}}$:
\begin{equation}
\delta {\mathcal {H}} = \left(
\frac {2\alpha _{0}^{2} \omega _{0}^{3}}{r^{2} \alpha _{0}'}
+ \mu _{0} \alpha _{0}'\omega _{0}
\right) \delta {\mathcal {G}}.
\label{eq:deltaH}
\end{equation}
Substituting for $\delta {\mathcal {H}}$ from (\ref{eq:deltaH}) into the second constraint equation (\ref{eq:constraint2}) then gives a first order differential equation for $\delta {\mathcal {G}}$:
\begin{equation}
0 = \delta {\mathcal {G}}' + \left( \frac {2}{r} - \frac {\mu _{0}'}{\mu _{0}} -\frac {3S_{0}'}{S_{0}} + \frac {3\alpha _{0}'}{\alpha _{0}}
+ \frac {2\omega _{0}'}{\omega _{0}} - \frac {2\alpha _{0}\omega _{0}^{2}}{r^{2} \mu _{0} \alpha _{0}'} \right) \delta {\mathcal {G}},
\end{equation}
which can be readily integrated to give
\begin{equation}
\delta {\mathcal {G}} (r) =
\frac { {\mathcal {G}}_{0} \mu _{0}S_{0}^{3} e^{{\mathcal {I}}} }{r^{2} \alpha _{0}^{3} \omega _{0}^{2}} ,
\label{eq:deltaG}
\end{equation}
where ${\mathcal {G}}_{0}$ is an arbitrary constant and
\begin{equation}
{\mathcal {I}} = \int _{r'=r_{0}}^{r} \frac {2\alpha _{0}(r') \omega _{0}(r')^{2}}{r'^{2} \mu _{0}(r') \alpha _{0}'(r')} dr' .
\label{eq:Iint}
\end{equation}
The lower limit $r_{0}$ in (\ref{eq:Iint}), and the consequences for $\delta {\mathcal {G}}$, depend on whether we are considering soliton or black hole solutions.

For black hole solutions, the expansions (\ref{eq:horizon}) mean that the integrand in (\ref{eq:Iint}) is regular as $r\rightarrow r_{h}$.
In this case we set the lower limit of integration $r_{0}$ to be $r_{h}$ and then ${\mathcal {I}} = O(r-r_{h})$ as $r\rightarrow r_{h}$.
At the black hole event horizon, we require $\delta \mu =0$ and $\delta \omega $ to be finite: we note by Lemma 1 that this statement is gauge invariant.
Therefore, from (\ref{eq:deltamu}), recalling that we have already set $\delta {\mathcal {F}}\equiv 0$, it must be the case that $(r-r_h)\delta \kappa \rightarrow 0$ as $r\rightarrow r_h $, so that both $\delta \kappa $ and $(r-r_{h})\delta \kappa'$ are integrable at the horizon.
We also require the perturbations $\delta S $, $\delta \psi $ and $\delta \xi $ to be finite at the event horizon.
While these quantities change under an infinitesimal co-ordinate transformation generated by (\ref{eq:gauge-freedom}), from (\ref{eq:gf-del-s}, \ref{eq:gf-del-xi}) we can see that such a transformation maintains the finiteness of these perturbations at the horizon.
Considering the first YM perturbation equation (\ref{eq:YMpert1}) as an ODE in $r$ for $\delta\psi = \delta \xi'-\delta{\dot{\eta}}$, carrying out a formal integration, and imposing the condition that $\delta\psi$ remains finite at the horizon, we find that $(r-r_{h})^{-1}\delta\xi$ must be integrable at $r=r_h$.
Therefore, from (\ref{eq:deltaS}), we see that $\delta {\mathcal {G}}$ must also be integrable at the horizon.
However, from the definition (\ref{eq:deltaG}), the properties of ${\mathcal {I}}$ as $r\rightarrow r_{h}$ and the boundary conditions (\ref{eq:horizon}), we see that
\begin{equation}
\delta {\mathcal {G}} = \frac {2{\mathcal {G}}_{0} S_{h}^{3}}{\left( r- r_{h} \right) r_{h}^{4} \alpha _{h}'^{3}} + O(1)
\end{equation}
as $r\rightarrow r_{h} $.
Thus $\delta{\mathcal {G}}$ fails to be integrable at the horizon unless ${\mathcal {G}}_{0}=0$. Therefore both $\delta {\mathcal {G}}$ and $\delta {\mathcal {H}}$ must vanish identically in the black hole case.

For soliton solutions, from the expansions (\ref{eq:origin}) we see that the integrand in (\ref{eq:Iint}) is $O(r^{-1})$ as $r\rightarrow 0$.
In this case we choose the lower limit of integration to be $r_{0}=1$.  With this choice, as $r\rightarrow 0$ we have
${\mathcal {I}} = 2\log r + O(1)$ and therefore $e^{\mathcal {I}} = O(r^{2})$ as $r\rightarrow 0$.
Substituting this and the expansions (\ref{eq:origin}) into (\ref{eq:deltaG}), it can be seen that $\delta {\mathcal {G}} = O(r^{-3})$ as $r\rightarrow 0$
unless ${\mathcal {G}}_{0}=0$.
In order to keep the origin regular, the perturbation $\delta \omega $ must remain finite at $r=0$, and we must also have $\delta \mu \rightarrow 0$ as $r\rightarrow 0$ to avoid a curvature singularity. As in the black hole case, these are diffeomorphism-invariant statements.
From (\ref{eq:deltamu}) and the boundary conditions (\ref{eq:origin}) we see that $\delta \kappa $ also remains finite as $r\rightarrow 0$, and, as a consequence, $r\delta \kappa '\rightarrow 0$ as $r\rightarrow 0$.
We also require $\delta S$, $\delta \psi $ and $\delta \xi $ to be finite at the origin.  As in the black hole case, using the results of Lemma 1, this requirement does not change if an infinitesimal co-ordinate transformation generated by (\ref{eq:gauge-freedom}) is applied.
With this assumption and the above behaviour of $\delta \kappa $ and its derivative as $r\rightarrow 0$, equation (\ref{eq:deltaS}) then implies that, at worst, $r^{2}\delta {\mathcal {G}}\rightarrow 0$ as $r\rightarrow 0$.
Therefore, as in the black hole case, the only possibility is ${\mathcal {G}}_{0}=0$, so that both $\delta {\mathcal {G}}$ and $\delta {\mathcal {H}}$
vanish identically.​

Setting $\delta {\mathcal {F}}\equiv 0$, $\delta {\mathcal {G}} \equiv 0$ and $\delta {\mathcal {H}}\equiv 0$,  the perturbation equations (\ref{eq:pair1}, \ref{eq:pair2}) simplify to
\numparts
\begin{eqnarray}
-\delta {\ddot {\kappa }} & = &
-\partial _{r_{*}}^{2} \delta \kappa
+ 2\alpha _{0} \partial _{r_{*}} \delta \omega
+ {\mathcal {E}}_{1} \delta \kappa + {\mathcal {E}}_{2} \delta \omega ,
\label{eq:pair1final}
\\
-\delta {\ddot {\omega }} & = &
- \partial _{r_{*}}^{2} \delta \omega
- 2\alpha _{0} \partial _{r_{*}} \delta \kappa
+ {\mathcal {E}}_{3} \delta \kappa + {\mathcal {E}}_{4} \delta \omega ,
\label{eq:pair2final}
\end{eqnarray}
\endnumparts
which do not involve the perturbation $\delta \xi $.
Once perturbations $\delta \kappa $, $\delta \omega $ solving (\ref{eq:pair1final}, \ref{eq:pair2final}) have been found, the perturbation $\delta \xi $
is computed by solving the constraint equation (\ref{eq:deltaxi}), which now simplifies to
\begin{eqnarray}
\left( \frac {1}{\alpha _{0}}\delta \xi \right) '
& = &
{\mathfrak {F}}(\delta \kappa, \delta \omega )
\nonumber \\ & = &
\frac {\mu _{0}S_{0}}{\alpha _{0}\omega _{0}} \delta \kappa ''
- \frac {2}{\omega _{0}} \delta \omega '
- \frac {\mu _{0}S_{0}}{\alpha _{0}\omega _{0}} \left(
\frac {\alpha _{0}'}{\alpha _{0}} - \frac {\mu _{0}'}{\mu _{0}}  - \frac {S_{0}'}{S_{0}}
\right) \delta \kappa '
\nonumber \\ & &
+\frac{ 4\omega _{0}' }{r} \left(
\frac {\omega _{0}\left( 1 - \omega _{0}^{2} \right) }{r^{2}\mu _{0}\omega _{0}'}
+ \frac {1}{r}
+ \frac {r}{2\omega _{0}^{2}}
+ \frac {\mu _{0}'}{\mu _{0}}
+ \frac {S_{0}'}{S_{0}}
 \right) \delta \omega
\nonumber \\ & &
- \frac {4\alpha _{0}\omega _{0}}{r\mu _{0}S_{0}} \left(
\frac {r}{4\omega _{0}^{2}}
+ \frac {\mu_{0}S_{0}^{2}\left( 1- \omega _{0}^{2} \right)}{4r\alpha _{0}^{2} \omega _{0}^{2}}
- \frac {1}{r}
+ \frac {\alpha _{0}'}{\alpha _{0}}
- \frac {\mu _{0}'}{\mu _{0}} -\frac {S_{0}'}{S_{0}}
\right. \nonumber \\ & & \left.
+ \frac {r\mu _{0}^{2}S_{0}^{2} \alpha _{0}'\omega _{0}'}{4\alpha _{0}^{3}\omega _{0}^{3}}
+ \frac {r\mu _{0}^{2}S_{0}^{2} \omega _{0}'^{2}}{2\alpha _{0}^{2}\omega _{0}^{4}}
 \right) \delta \kappa ,
\label{eq:deltaxifinal}
\end{eqnarray}
where we have defined a new quantity ${\mathfrak {F}}(\delta \kappa, \delta \omega )$ which depends on $\delta \kappa $, $\delta \omega $ and the equilibrium solutions.

The fact that we do not have a dynamical equation for $\delta \xi $ can be understood from the analysis of section \ref{sec:gauge}.
As discussed in that section, we have a residual infinitesimal diffeomorphism gauge freedom generated by (\ref{eq:gauge-freedom}), which corresponds to a redefinition of the time co-ordinate.  Such a gauge transformation changes $\delta \xi $ according to (\ref{eq:gf-del-xi}). Since this is a gauge transformation, the perturbation equations should be independent of the choice of the gauge function $x(t)$.  When $\delta \xi $ changes as in (\ref{eq:gf-del-xi}), the quantity $\left( \alpha _{0}^{-1}\delta \xi \right) ' $ does not change, and so the equation (\ref{eq:deltaxifinal}) remains invariant.  If we had a dynamical equation for $\delta \xi $ involving time derivatives of $\delta \xi $, such an equation would not be invariant under the gauge transformation (\ref{eq:gf-del-xi}). Since we have already eliminated $\delta \psi $, the only other perturbation which changes under the residual diffeomorphism gauge transformation is $\delta S$ (\ref{eq:gf-del-s}). However, the changes in $\delta \xi $ (\ref{eq:gf-del-xi})
and $\delta S $ (\ref{eq:gf-del-s}) mean that the equation (\ref{eq:deltaS}) for $\delta S$ remains invariant under the gauge transformation.

Our purpose in this paper is to prove the existence of dyonic equilibrium soliton and black hole solutions of the static EYM field equations which are stable under linear perturbations satisfying (\ref{eq:pair1final}, \ref{eq:pair2final}, \ref{eq:deltaxifinal}). We turn to this proof in the next section.

\section{Existence of stable dyonic solitons and black holes}
\label{sec:stable}

\subsection{General argument}
\label{sec:stabgen}

The perturbation equations (\ref{eq:pair1final}, \ref{eq:pair2final}) can be written in the compact form
\begin{equation}
-{\ddot {{\bmath {v}}}} = -\partial _{r_{*}}^{2} {\bmath {v}} + {\mathcal {D}} \partial _{r_{*}} {\bmath {v}} + {\mathcal {E}} {\bmath {v}}
= {\mathcal {U}} {\bmath {v}},
\label{eq:operator}
\end{equation}
where ${\bmath {v}} = \left( \delta \kappa , \delta \omega \right) ^{T}$ is the vector of perturbations, and the matrices ${\mathcal {D}}$ and
${\mathcal {E}}$ are given by
\begin{equation}
{\mathcal {D}} = \left(
\begin{array}{cc}
0 & 2\alpha _{0} \\
-2\alpha _{0} &  0
\end{array}
\right) ,
\qquad
{\mathcal {E}} = \left(
\begin{array}{cc}
{\mathcal {E}}_{1} & {\mathcal {E}}_{2} \\
{\mathcal {E}}_{3} & {\mathcal {E}}_{4}
\end{array}
\right) ,
\label{eq:DandE}
\end{equation}
with ${\mathcal {E}}_{1},\ldots ,{\mathcal {E}}_{4}$ given in (\ref{eq:E1}--\ref{eq:E4}).
We restrict our attention to static equilibrium solutions for which the magnetic gauge field function $\omega _{0}(r)$ has no zeros.
In this case the functions  ${\mathcal {E}}_{1},\ldots ,{\mathcal {E}}_{4}$ are regular for all $r>0$ in the soliton case and all $r\ge r_{h}$ in the black hole case.
They diverge as $r\rightarrow 0$ for soliton equilibrium solutions, but this is not an issue, as discussed after (\ref{eq:eigenvalue}) below.

In (\ref{eq:operator}) we have defined an operator ${\mathcal {U}}$ by
\begin{equation}
{\mathcal {U}} = -\partial _{r_{*}}^{2} + {\mathcal {D}} \partial _{r_{*}} + {\mathcal {E}}.
\end{equation}
The operator ${\mathcal {U}}$ will be symmetric if
\begin{equation}
{\mathcal {D}}^{T} = -{\mathcal {D}} \qquad {\mbox {and}} \qquad {\mathcal {E}}^{T} = {\mathcal {E}} - \partial _{r_{*}} {\mathcal {D}}.
\end{equation}
The first of these conditions is clearly satisfied.
With the form of ${\mathcal {D}}$ in (\ref{eq:DandE}), the second condition is satisfied if
\begin{equation}
{\mathcal {E}}_{2} - {\mathcal {E}}_{3} = \partial _{r_{*}} \left( 2 \alpha _{0} \right) = 2\mu _{0} S_{0} \alpha _{0}' ,
\end{equation}
which can be seen to be the case from (\ref{eq:E3}).
Therefore ${\mathcal {U}}$ is a symmetric operator.

To derive sufficient conditions for ${\mathcal {U}}$ to be a positive operator, we first write it in the form
\begin{equation}
{\mathcal {U}} = \chi ^{\dagger }\chi + {\mathcal {V}},
\label{eq:Udef}
\end{equation}
where the operator ${\mathcal {V}}$ does not contain any derivatives,
\begin{equation}
\chi = \partial _{r_{*}} - {\mathcal {Z}}, \label{eq:chi-def}
\end{equation}
and ${\mathcal {Z}}$ is a $2\times 2$ matrix with entries
\begin{equation}
{\mathcal {Z}} = \left(
\begin{array}{cc}
0 & {\mathcal {Z}}_{12} \\
{\mathcal {Z}}_{21} &  0
\end{array}
\right) .
\end{equation}
We choose the functions ${\mathcal {Z}}_{12}$ and ${\mathcal {Z}}_{21}$ such that
\begin{equation}
\partial _{r_{*}} {\mathcal {Z}}_{12} = {\mathcal {E}}_{2}, \qquad
\partial _{r_{*}} {\mathcal {Z}}_{21} = {\mathcal {E}}_{3}.
\label{eq:Zdef}
\end{equation}
Then the matrix ${\mathcal {V}}$ takes the form
\begin{equation}
{\mathcal {V}} = \left(
\begin{array}{cc}
{\mathcal {E}}_{1} - {\mathcal {Z}}_{21}^{2} & 0 \\
0 & {\mathcal {E}}_{4} - {\mathcal {Z}}_{12}^{2}
\end{array}
\right) .
\label{eq:Vform}
\end{equation}
Therefore the operator ${\mathcal {U}}$ will be a positive symmetric operator if the functions ${\mathcal {K}}_{1}$ and ${\mathcal {K}}_{2}$ are positive
everywhere, where
\begin{equation}
{\mathcal {K}}_{1} = {\mathcal {E}}_{1} - {\mathcal {Z}}_{21}^{2}, \qquad
{\mathcal {K}}_{2} = {\mathcal {E}}_{4} - {\mathcal {Z}}_{12}^{2}.
\label{eq:Kdef}
\end{equation}
If ${\mathcal{U}}$ is a positive symmetric operator then the static equilibrium solutions will be stable under linear, spherically symmetric perturbations.
It turns out that there are three related stability properties: these will be described in detail in section \ref{sec:stabdetails}.

\begin{figure}
\begin{center}
\includegraphics[width=10cm]{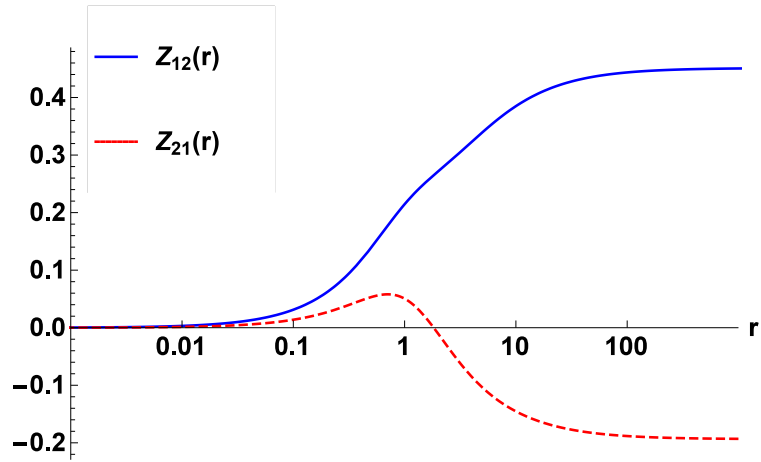}
\end{center}
\caption{Functions ${\mathcal {Z}}_{12}$ (blue, solid) and ${\mathcal {Z}}_{21}$ (red, dotted) defined by (\ref{eq:Zdef}) for a typical nodeless dyonic soliton solution with $\ell =1$, $\alpha _{1}=0.08715$ and $\omega _{2}=0.2$.
The additive constant of integration has been chosen so that both functions vanish at the origin.
For this example, ${\mathcal {Z}}_{12}$ is monotonically increasing with $r$; on the other hand ${\mathcal {Z}}_{21}$ has a maximum and is negative for sufficiently large $r$.
}
\label{fig:three}
\end{figure}

\begin{figure}
\begin{center}
\includegraphics[width=10cm]{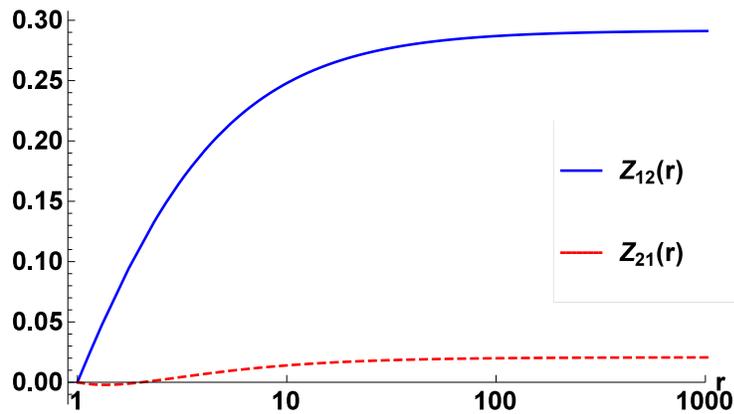}
\end{center}
\caption{Functions ${\mathcal {Z}}_{12}$ (blue, solid) and ${\mathcal {Z}}_{21}$ (red, dotted) defined by (\ref{eq:Zdef}) for a typical nodeless dyonic black hole solution with $\ell =1$, $r_{h}=1$, $\alpha _{h}'=0.09974$ and $\omega _{h}=0.9$. The additive constant of integration has been chosen so that both functions vanish on the horizon. As in the example shown in figure \ref{fig:three}, in this case ${\mathcal {Z}}_{12}$ is monotonically increasing as $r$ increases.  The other function, ${\mathcal {Z}}_{21}$, has a minimum just outside the event horizon and then slowly increases as $r$ increases.}
\label{fig:four}
\end{figure}

For the moment, let us assume that to prove the stability (as will be characterized in section \ref{sec:stabdetails}) of the static equilibrium solutions it is sufficient to prove that the functions ${\mathcal {K}}_{1}$, ${\mathcal {K}}_{2}$ (\ref{eq:Kdef}) are positive everywhere and, in addition, have the asymptotic and other properties used in the stability arguments in section \ref{sec:stabdetails}.
Before proving the existence of dyonic solutions for which this is the case, we present a couple of numerical examples, considering the static equilibrium solutions shown in figures \ref{fig:one} and \ref{fig:two}.
Having found an equilibrium solution by integrating the static field equations (\ref{eq:static1}--\ref{eq:static4}), the first step in showing that they are stable is to numerically integrate (\ref{eq:Zdef}) to find the functions ${\mathcal {Z}}_{12}$ and ${\mathcal {Z}}_{21}$.
In figures \ref{fig:three} and \ref{fig:four} we show the results of these integrations for the equilibrium solutions presented in figures \ref{fig:one}
and \ref{fig:two} respectively.
Note that the equations (\ref{eq:Zdef}) only define the functions ${\mathcal {Z}}_{12}$ and ${\mathcal {Z}}_{21}$ up to the addition of an arbitrary constant. This constant is chosen so that the functions vanish at either the origin, or the black hole event horizon, as applicable.

\begin{figure}
\begin{center}
\includegraphics[width=10cm]{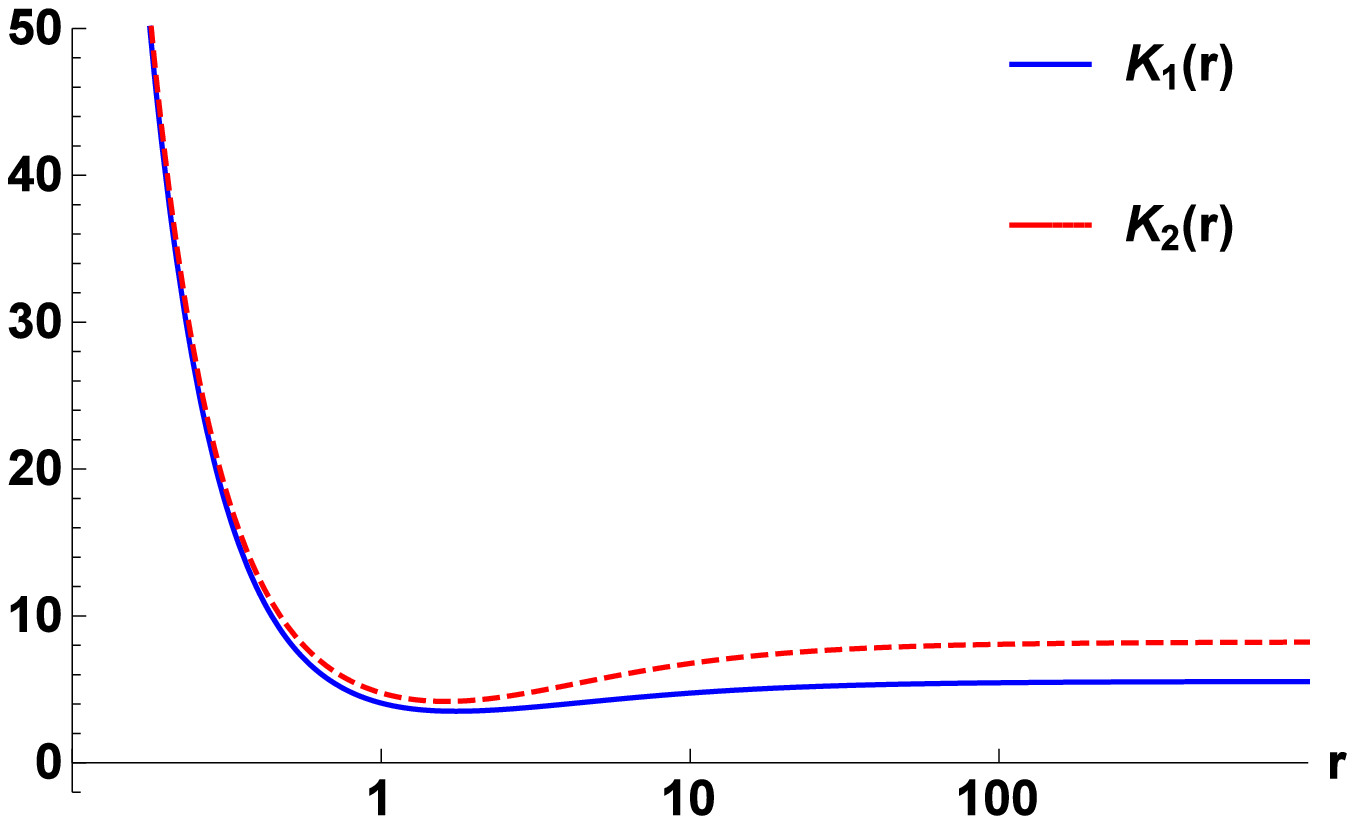}
\end{center}
\caption{Functions ${\mathcal {K}}_{1}$ (blue, solid) and ${\mathcal {K}}_{2}$ (red, dotted) defined by (\ref{eq:Kdef}) for a typical nodeless dyonic soliton solution with $\ell =1$, $\alpha _{1}=0.08715$ and $\omega _{2}=0.2$. Both functions are positive for all $r$.  For small $r$, they exhibit very similar behaviour, only becoming distinguishable for $r\approx 1$.  As $r\rightarrow 0$, both functions diverge, in accordance with (\ref{eq:ki-soliton-origin}). They both have a minimum, and then tend to (positive) constants as $r\rightarrow \infty $.
}
\label{fig:five}
\end{figure}

\begin{figure}
\begin{center}
\includegraphics[width=10cm]{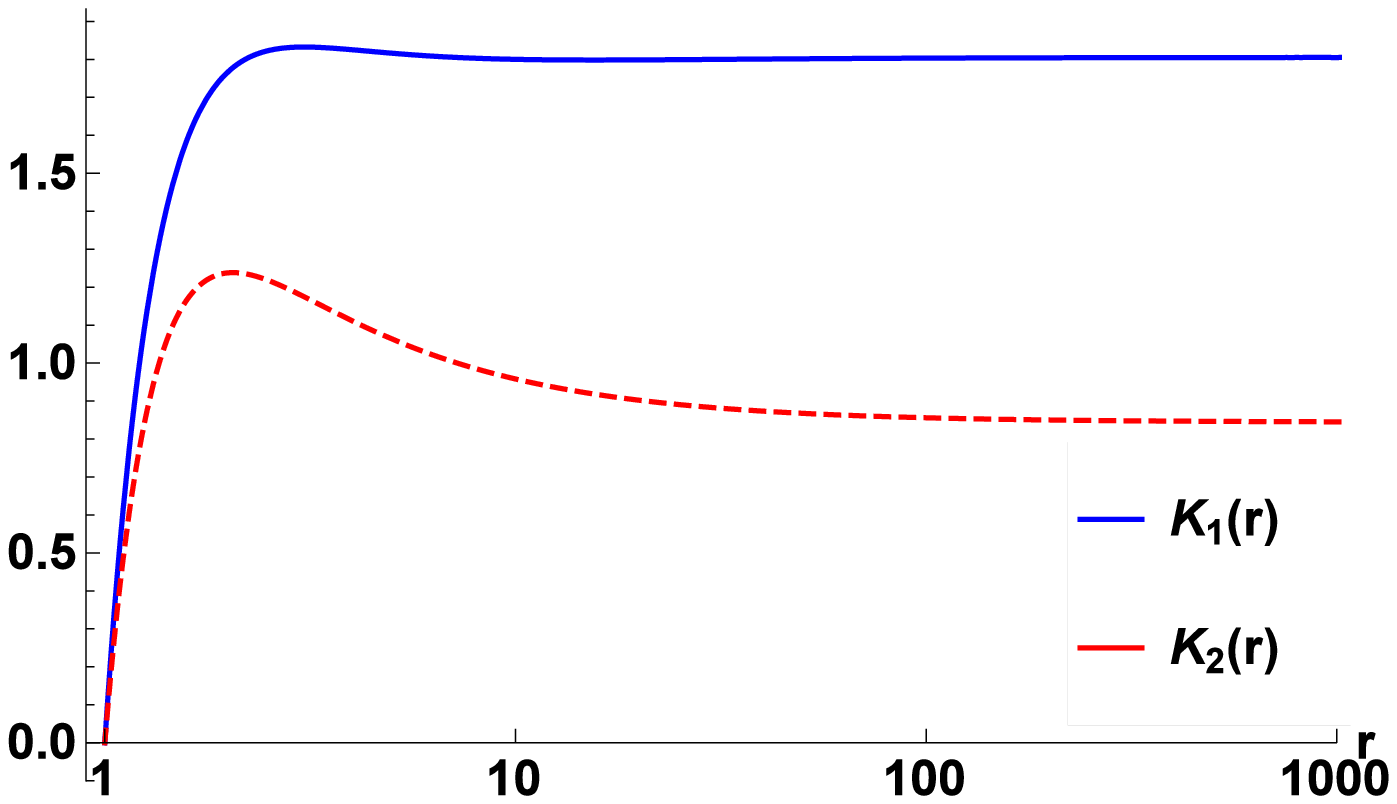}
\end{center}
\caption{Functions ${\mathcal {K}}_{1}$ (blue, solid) and ${\mathcal {K}}_{2}$ (red, dotted) defined by (\ref{eq:Kdef}) for a typical nodeless dyonic black hole solution with $\ell =1$, $r_{h}=1$, $\alpha _{h}'=0.09974$ and $\omega _{h}=0.9$.
Both functions are positive for all $r\ge r_{h}$. Both vanish on the horizon, then rapidly increase to maximum values, before tending to (positive) constants as $r\rightarrow \infty $.
}
\label{fig:six}
\end{figure}

Once we have the functions ${\mathcal {Z}}_{12}$ and ${\mathcal {Z}}_{21}$, we then compute the quantities ${\mathcal {K}}_{1}$ and ${\mathcal {K}}_{2}$ (\ref{eq:Kdef}). If we can show that ${\mathcal {K}}_{1}$ and ${\mathcal {K}}_{2}$ are positive for all $r$, then the operator ${\mathcal {U}}$ (\ref{eq:Udef}) is a positive symmetric operator and hence, as argued above, the equilibrium solutions are stable. In
figures \ref{fig:five} and \ref{fig:six} respectively we present the results of calculating ${\mathcal {K}}_{1}$ and ${\mathcal {K}}_{2}$ for the dyonic soliton and black hole solutions shown in figures \ref{fig:one} and \ref{fig:two}. In both figures \ref{fig:five} and \ref{fig:six}, it can be seen that the two functions ${\mathcal {K}}_{1}$ and ${\mathcal {K}}_{2}$ are positive everywhere.

We now turn to the proof of the existence of dyonic soliton and black hole solutions which are stable under linear, spherically symmetric, perturbations of the metric and gauge field. The stability criterion that we apply is that the operator ${\mathcal{U}}$ governing the evolution of the perturbations (\ref{eq:Udef}) is a symmetric, positive operator. As we have seen, this condition reduces to establishing that the functions ${\mathcal{K}}_1$ and ${\mathcal{K}}_2$ are positive everywhere - that is, for all $r>r_h$ in the black hole case, and for all $r>0$ in the soliton case. We establish this in each of these cases by recalling some of the results from \cite{Nolan}, in which we proved the existence of the background space-times being studied here. We consider the black hole case first.

\subsection{Existence of stable dyonic black holes}
\label{sec:stabBH}

In \cite{Nolan}, we proved the existence of static, spherically symmetric, asymptotically adS black hole solutions of the $\mathfrak{su}(2)$ EYM equations for which the metric and gauge field functions $\mu_0, S_0, \alpha_0$ and $\omega_0$ have the following properties:
\begin{itemize}
\item[(i)] $(\mu_0, S_0, \alpha_0,\omega_0) \in C^1([r_h,+\infty),\mathbb{R}^2)\times C^2([r_h,+\infty),\mathbb{R}^2)$;
\item[(ii)] $\mu_0(r_h)=0$ and $\mu_0(r)>0$, $\omega_0(r)>0$ and $S_0(r)>0$ for all $r>r_h$;
\item[(iii)] there is a continuous mapping of the initial data (imposed at the horizon) $(S_h,\omega_h,\alpha_h')\in\mathbb{R}^3$ to the solution $(\mu_0, S_0, \alpha_0,\omega_0) \in C^1([r_h,+\infty),\mathbb{R}^2)\times C^2([r_h,+\infty),\mathbb{R}^2)$;
\item[(iv)] this continuous mapping yields a unique global solution of the field equations in a neighbourhood of the trivial datum $(S_h,\omega_h,\alpha_h')=(1,1,0)\in\mathbb{R}^3$ which gives rise to Schwarzschild-adS space-time as an embedded solution of the system. We will refer to this as the \textit{trivial solution}: it nevertheless plays an important role in what follows.
\end{itemize}

The trivial solution is characterized by (\ref{eq:SadS}), so that
\begin{equation}
\mu_{0}(r) = 1-\frac{2M}{r}-\frac{\Lambda}{3}r^2.
\label{eq:trivial}
\end{equation}
A simple calculation using (\ref{eq:E1}--\ref{eq:E4}) then yields ${\mathcal{E}}_2={\mathcal{E}}_3=0$ and
\begin{equation}
{\mathcal{E}}_1 = {\mathcal{E}}_4 = \frac{2}{r^2}\left( 1-\frac{2M}{r}-\frac{\Lambda}{3}r^2 \right).
\label{eq:E1-E4-trivial}
\end{equation}
By a choice of constant of integration, we can then take ${\mathcal{Z}}_{12}={\mathcal{Z}}_{21}=0$, and so
\begin{equation}
{\mathcal{K}}_1 = {\mathcal{K}}_2 = \frac{2}{r^2}\left( 1-\frac{2M}{r}-\frac{\Lambda}{3}r^2 \right).
\label{eq:K1-K2-trivial}
\end{equation}
It is then immediate that both ${\mathcal{K}}_1$ and ${\mathcal{K}}_2$ are positive outside the horizon.

For the non-trivial background solutions, we can use the asymptotic forms (\ref{eq:horizon}) to establish that
\begin{equation}
{\mathcal {E}}_i(r) = O(r-r_h),\qquad r\rightarrow r_h, \qquad  i=1,2,3,4,
\label{eq:Ei-hor}
\end{equation}
and
\begin{equation}
{\mathcal {E}}_i(r) = {\mathcal {E}}_{i,\infty}+O(r^{-1}),\qquad r\rightarrow +\infty, \qquad  i=1,2,3,4.
\label{eq:Ei-inf}
\end{equation}
Then
\begin{equation}
\frac{{\mathcal {E}}_i(r)}{\mu_0S_0} =O(1),\qquad r\rightarrow r_h, \qquad i=2,3,
\label{eq:Ei-ratio}
\end{equation}
and so we can take
\begin{equation}
{\mathcal{Z}}_{12}(r) = \int_{r'=r_h}^r \frac{{\mathcal {E}}_2(r')}{\mu_0(r')S_0(r')}dr', \quad
{\mathcal{Z}}_{21}(r) = \int_{r'=r_h}^r \frac{{\mathcal {E}}_3(r')}{\mu_0(r')S_0(r')}dr',
\label{eq:z21-int}
\end{equation}
giving
\begin{equation}
{\mathcal{Z}}_{12}(r)=O(r-r_h),\qquad {\mathcal{Z}}_{21}(r)=O(r-r_h),\qquad r\rightarrow r_h ,
\label{eq:z12-asypm}
\end{equation}
and
\begin{equation}
{\mathcal{Z}}_{12}(r)=O(r^{-1}),\qquad {\mathcal{Z}}_{21}(r)=O(r^{-1}),\qquad r\rightarrow +\infty .
\label{eq:z12-inf}
\end{equation}
It follows that, as $r\rightarrow r_{h}$,
\begin{equation}
{\mathcal{K}}_1(r) = {\mathcal {E}}_1(r) + O([r-r_h]^2),
\qquad {\mathcal{K}}_2(r) = {\mathcal {E}}_4(r) + O([r-r_h]^2),
\label{eq:k2-asymp}
\end{equation}
and, at infinity,
\begin{equation}
{\mathcal{K}}_1(r) = {\mathcal {E}}_1(r) + O(r^{-2}),
\qquad {\mathcal{K}}_2(r) = {\mathcal {E}}_4(r) + O(r^{-2}), \qquad r\rightarrow +\infty.
\label{eq:k2-inf}
\end{equation}
Therefore ${\mathcal{K}}_i$, $i=1,2$ are continuous as $r\rightarrow r_h$. It follows from (\ref{eq:E1}--\ref{eq:E4}, \ref{eq:z21-int}) and from points (i) and (ii) in the list above that the ${\mathcal{K}}_i$ are continuous on $(r_h,+\infty)$. Thus
\begin{equation}
{\mathcal{K}}_i\in C^0[r_h,+\infty),\qquad i=1,2.
\label{ki-cts}
\end{equation}
Furthermore, it follows from this observation and from point (iii) above that for each $r_h>0$, the ${\mathcal{K}}_i$ depend continuously on the initial data $(S_h,\omega_h,\alpha_h')\in\mathbb{R}^3$. Thus we can conclude the following:

\begin{Proposition}
For each $\Lambda<0$ and $r_h>0$, there exists a neighbourhood $U$ of the trivial initial data point $(S_h,\omega_h,\alpha_h')=(1,1,0)\in\mathbb{R}^3$ such that for all $(S_h,\omega_h,\alpha_h')\in U$, the corresponding unique black hole solution of the static EYM equations, whose existence is guaranteed by Proposition 7 of \cite{Nolan}, satisfies
\begin{equation}
{\mathcal{K}}_i(r)>0 \hbox{ for all } r>r_h,\quad i=1,2,
\label{eq:ki-pos}
\end{equation}
and thus is linearly stable under spherically symmetric perturbations.
\end{Proposition}

From (\ref{eq:Ei-hor}, \ref{eq:z12-asypm}), the functions ${\mathcal {K}}_{i}$, $i=1,2$ vanish at the event horizon.
We note that a necessary condition for them to be positive outside the horizon is that
\begin{eqnarray}
k_1 &:=& \left. {\mathcal{K}}_1'(r)\right|_{r=r_h} = \left.{\mathcal{E}}_1'(r)\right|_{r=r_h} >0,
\nonumber \\
k_2 &:=& \left. {\mathcal{K}}_2'(r)\right|_{r=r_h} = \left.{\mathcal{E}}_4'(r)\right|_{r=r_h} >0.
\end{eqnarray}
Using the expressions (\ref{eq:E1}, \ref{eq:E4}), we find that
\begin{eqnarray}
k_1 &=& \frac{1}{r_h^5}\left\{ \alpha_h'^2r_h^4(3\omega_h^2-1)+S_h^2(1+\omega_h^2)\left[ r_h^2-\Lambda r_h^4-(1-\omega_h^2)^2\right] \right\},
\nonumber \\
k_2 &=& \frac{1}{r_h^5}\left\{ -\alpha_h'^2r_h^4(3\omega_h^2-1)+S_h^2\left[1-\omega_h^2+\omega_h^4-\omega_h^6 \right. \right.
\nonumber \\ & & \left. \left.
+r_h^2(1-\Lambda r_h^2)(3\omega_h^2-1)\right] \right\}.
\label{eq:k12bh}
\end{eqnarray}
Expanding (\ref{eq:k12bh}) about the trivial data shows how a neighbourhood of allowed values of the initial data for which $k_{i}$, $i=1,2$ are positive may arise:
\begin{eqnarray}
k_1 &=& \frac{2}{r_h^3}\left(1+3\frac{r_h^2}{\ell^2}\right)S_h^2 + O(\alpha_h')^2 + O(\omega_h-1), \nonumber \\
k_2 &=& \frac{2}{r_h^3}\left(1+3\frac{r_h^2}{\ell^2}\right)S_h^2 + O(\alpha_h')^2 + O(\omega_h-1).
\end{eqnarray}

\subsection{Existence of stable dyons}
\label{sec:stabsol}

Positivity of the ${\mathcal{K}}_i$ in the soliton case follows by a similar argument.

In \cite{Nolan}, we proved the existence of static, spherically symmetric, asymptotically adS, soliton solutions of the $\mathfrak{su}(2)$ EYM equations for which the metric and gauge field functions $\mu_0, S_0, \alpha_0$ and $\omega_0$ have the following properties:
\begin{itemize}
\item[(i)] $(\mu_0, S_0, \alpha_0,\omega_0) \in C^1([0,+\infty),\mathbb{R}^2)\times C^2([0,+\infty),\mathbb{R}^2)$;
\item[(ii)] $\mu_0(r)>0$, $\omega_0(r)>0$ and $S_0(r)>0$ for all $r\geq 0$ and these functions satisfy the asymptotic behaviour of (\ref{eq:origin}) at the origin;
\item[(iii)] there is a continuous mapping of the initial data (imposed at the origin) $(S_1,\omega_2,\alpha_1)\in\mathbb{R}^3$ to the solution $(\mu_0, S_0, \alpha_0,\omega_0) \in C^1([0,+\infty),\mathbb{R}^2)\times C^2([0,+\infty),\mathbb{R}^2)$;
\item[(iv)] this continuous mapping yields a unique global solution of the field equations in a neighbourhood of the trivial datum $(S_1,\omega_2,\alpha_1)=(1,0,0)\in\mathbb{R}^3$ which gives rise to adS space-time as an embedded solution of the system. This is the trivial solution in the solitonic case.
\end{itemize}

For the trivial solution, we find ${\mathcal {E}}_2={\mathcal {E}}_3=0$ and
\begin{equation}
{\mathcal {E}}_1(r) =  {\mathcal {E}}_4(r) = \frac{2}{r^2}\left(1+\frac{r^2}{\ell^2}\right).
\label{eq-ei-ads}
\end{equation}
Then as in the black hole case, we can take ${\mathcal{Z}}_{12}={\mathcal{Z}}_{21}=0$ in the trivial background, and positivity of ${\mathcal{K}}_1={\mathcal {E}}_1$ and ${\mathcal{K}}_2={\mathcal {E}}_4$ on $(0,+\infty)$ is immediate.

For the non-trivial background space-times, we can use (\ref{eq:origin}, \ref{eq:E2}, \ref{eq:E3}) to establish that
\begin{equation}
{\mathcal {E}}_2(r) = 2\alpha_1S_1(1+4\omega_2) + O(r), \qquad
{\mathcal {E}}_3(r) = 8\alpha_1S_1\omega_2 + O(r),
\label{eq:e3-asymp-origin}
\end{equation}
as $r\rightarrow 0$.
The general form of the functions ${\mathcal {E}}_2$ (\ref{eq:E2}), ${\mathcal {E}}_3$ (\ref{eq:E3}) and the properties of the background metric functions listed above, along with the limiting behaviour at the origin, then shows that ${\mathcal {E}}_2$ and ${\mathcal {E}}_3$ are continuous on $[0,+\infty)$. The same is true of ${\mathcal {E}}_i/(\mu_0S_0)$ for $i=2,3$ (again appealing to (\ref{eq:origin}) and the properties listed above) and so, from (\ref{eq:Zdef}) we can take
\begin{equation}
{\mathcal{Z}}_{12}(r) = \int_{r'=0}^r \frac{{\mathcal {E}}_2(r')}{\mu_0(r')S_0(r')}dr' \in C^1[0,+\infty),
\label{eq:z12-soliton}
\end{equation}
which yields
\begin{equation}
{\mathcal{Z}}_{12}(r) =2\alpha_1(1+4\omega_2)r + O(r^2),\qquad r\rightarrow 0.
\label{eq:z12-soliton-limit}
\end{equation}
Likewise,
\begin{equation}
{\mathcal{Z}}_{21}(r) = \int_{r'=0}^r \frac{{\mathcal {E}}_3(r')}{\mu_0(r')S_0(r')}dr' \in C^1[0,+\infty),
\label{eq:z21-soliton}
\end{equation}
with
\begin{equation}
{\mathcal{Z}}_{21}(r) =8\alpha_1\omega_2r + O(r^2),\qquad r\rightarrow 0.
\label{eq:z21-soliton-limit}
\end{equation}
We note also that the asymptotic behaviour at infinity as described by (\ref{eq:infinity}) yields
\begin{equation}
{\mathcal{Z}}_{12} = O(r^{-1}),\qquad {\mathcal{Z}}_{21}=O(r^{-1}),\qquad r\rightarrow+\infty. \label{z12-z21-infinity}
\end{equation}
At the origin, ${\mathcal{Z}}_{12}$ and ${\mathcal{Z}}_{21}$ are dominated by ${\mathcal {E}}_1$, ${\mathcal {E}}_4$, and we find that
\begin{equation}
{\mathcal{K}}_1 =\frac{2S_1^2}{r^2} + O(1),\qquad
{\mathcal{K}}_2 = \frac{2S_1^2}{r^2} + O(1),\qquad r\rightarrow 0,
\label{eq:ki-soliton-origin}
\end{equation}
and so the ${\mathcal{K}}_i$, $i=1,2$ are clearly positive in a (one-sided) neighbourhood of the origin.
At infinity, we find
\begin{equation}
{\mathcal{K}}_1 = O(1),\qquad {\mathcal{K}}_2 = O(1),\qquad r\rightarrow+\infty.
\label{eq:ki-soliton-infinity}
\end{equation}

With these details in place, positivity of the ${\mathcal{K}}_i$, $i=1,2$ on $(0,+\infty)$ follows in this case in exactly the same way as the black hole case, giving:
\begin{Proposition}
For each $\Lambda<0$, there exists a neighbourhood $U$ of the trivial initial data point $(S_1,\omega_2,\alpha_1)=(1,0,0)\in\mathbb{R}^3$ such that for all $(S_1,\omega_2,\alpha_1)\in U$, the corresponding unique soliton solution of the static EYM equations, whose existence is guaranteed by Proposition 9 of \cite{Nolan}, satisfies
\begin{equation}
{\mathcal{K}}_i(r)>0 \hbox{ for all } r>0,\quad i=1,2,
\label{eq:ki-pos-sol}
\end{equation}
and thus is linearly stable under spherically symmetric perturbations.
\end{Proposition}

\subsection{Nature of the stability}
\label{sec:stabdetails}

In the previous section we have proven the existence of dyonic soliton and black hole solutions of ${\mathfrak {su}}(2)$ EYM such that the functions ${\mathcal {K}}_{1}$ and ${\mathcal {K}}_{2}$ (\ref{eq:Kdef}) are positive everywhere.
Therefore the operator ${\mathcal {U}}$ (\ref{eq:Udef}) is a positive symmetric operator.
In this case we have three related stability properties, which we now discuss in detail.

\subsubsection{Mode stability.}
First, we consider time-periodic perturbations with frequency $\sigma $, so that
\begin{equation}
\delta \kappa (t,r) = e^{i\sigma t} \delta \kappa (r), \qquad
\delta \omega (t,r) = e^{i\sigma t} \delta \omega (r),
\label{eq:periodic}
\end{equation}
(we return to $\delta \xi $ shortly).
Then the perturbation equations (\ref{eq:operator}) for ${\bmath {v}}=(\delta \kappa (r), \delta \omega (r))^{T}$ take the form of a standard Schr\"odinger-like eigenvalue problem:
\begin{equation}
{\mathcal {U}} {\bmath {v}} = \sigma ^{2} {\bmath {v}}.
\label{eq:eigenvalue}
\end{equation}
To establish that all the eigenvalues $\sigma ^{2}$ must be positive, we need to impose suitable boundary conditions on the perturbations.
We set $\delta \kappa (r)$ and $\delta \omega (r)$ to vanish at the origin (for soliton solutions), event horizon (for black hole solutions) and at infinity.
With these conditions on $\delta \kappa (r)$ and $\delta \omega (r)$, the inner products $\langle {\bmath {v}}, {\bmath {v}} \rangle$ and
$\langle {\bmath {v}}, {\mathcal {U}} {\bmath {v}} \rangle $ are finite.
As usual, the inner products are defined by integrals over the range of the tortoise co-ordinate $r_{*}$: for solitons this is $r_{*} \in [0, r_{*{\mathrm {max}}}]$ for some constant $r_{*{\mathrm {max}}}$ while for black holes it is $r_{*}\in (-\infty, 0]$.
The only issue that may be of concern is the form of the zero-order terms in (\ref{eq:operator}) in the soliton case, which diverge as $r^{-2}$ at the origin (\ref{eq:ki-soliton-origin}). However the coefficient of $r^{-2}$ is sufficiently large and positive to ensure that solutions vanishing at the origin decay rapidly enough to yield integrable terms in the inner products.

With suitable boundary conditions in place, the positivity of the operator ${\mathcal {U}}$ implies that all the eigenvalues $\sigma ^{2}$ are positive, so that $\sigma $ is real and the perturbations do not grow exponentially with time.
There is one further remaining subtlety, namely that we do not have a dynamical equation for $\delta \xi (t,r)$.
Suppose however that we have perturbations $\delta \kappa (t,r)$, $\delta \omega (t,r)$ of the form (\ref{eq:periodic}) which satisfy
(\ref{eq:eigenvalue}) with $\sigma ^{2}>0$.
Then, by integrating (\ref{eq:deltaxifinal}), we find
\begin{equation}
\delta \xi (t,r) = \alpha _{0}(r) X(t) + \alpha _{0}(r) \int _{r'=r_{0}}^{r}  {\mathfrak {F}}(\delta \kappa, \delta \omega ) \, dr' ,
\label{eq:deltaxiint}
\end{equation}
where $X(t)$ is an arbitrary function of time $t$ and the quantity ${\mathfrak {F}}(\delta \kappa, \delta \omega )$ depends linearly on the perturbations $(\delta \kappa, \delta \omega )$ and can be found in full in (\ref{eq:deltaxifinal}).
Since the perturbations $(\delta \kappa, \delta \omega )$ are time-periodic (\ref{eq:periodic}), and given the form of ${\mathfrak {F}}(\delta \kappa, \delta \omega )$ in (\ref{eq:deltaxifinal}), the second term in (\ref{eq:deltaxiint}) does not grow exponentially with time.
The first term corresponds to an infinitesimal diffeomorphism generated by (\ref{eq:gauge-freedom}), and by a suitable choice of $x(t)$ in (\ref{eq:gf-del-xi}) can be transformed away.
Therefore $\delta \xi (t,r)$ is also time-periodic with $\sigma ^{2}>0$ and does not grow exponentially with time.
The fact that the metric perturbations $\delta \mu $ and $\delta S$ also do not grow exponentially with time can be deduced from their explicit forms (\ref{eq:deltamu}, \ref{eq:deltaS}) in terms of the matter perturbations.

\subsubsection{Finite energy.}
Next, we note that the form (\ref{eq:Udef}) of the perturbation equations, with ${\mathcal{V}}$ positive, gives rise to a positive, conserved energy for the system. That is, defining
\begin{equation}
E[\bmath{v}](t) = \frac12\langle\dot{\bmath{v}},\dot{\bmath{v}}\rangle+\frac12\langle{\bmath{v}},{\mathcal{U}}{\bmath{v}}\rangle,
\label{eq:energy}
\end{equation}
we have
\begin{eqnarray}
E[\bmath{v}](t) &\geq& 0 \hbox{ for all } \bmath{v};
\nonumber  \\
&=& 0 \Leftrightarrow \bmath{v}=\bmath{0}. \label{eq:E-zero}
\end{eqnarray}
Positivity follows from the decomposition (\ref{eq:Udef}), the definition of the adjoint operator $\chi^\dagger$ and positivity of ${\mathcal{V}}$:
\begin{equation}
E[\bmath{v}](t) = \frac12\|\dot{\bmath{v}}\|^2+\frac12\|{\mathcal{\chi}}\bmath{v}\|^2 +\frac12\langle\bmath{v},{\mathcal{V}}\bmath{v}\rangle,
\label{eq:E-pos-form}
\end{equation}
where the norm is that associated with the inner product. We note that the factor $\frac12$ appearing in (\ref{eq:energy}) is conventional, to make the link with the conserved energy of a particle in classical mechanics. Conservation of the energy corresponds to the fact that
\begin{equation}
\frac{d}{dt}\left\{ E[\bmath{v}](t)\right\} = 0.
\label{eq-E-constant}
\end{equation}
The derivation of this equation is straightforward, relying on the symmetry of ${\mathcal{V}}$ in (\ref{eq:Vform}), and on the identity
\begin{equation}
\langle\dot{\bmath{v}},\chi^\dagger\chi\bmath{v}\rangle = \langle\chi\dot{\bmath{v}},\chi\bmath{v}\rangle = \frac12\left(\langle\chi\dot{\bmath{v}},\chi\bmath{v}\rangle+\langle\chi{\bmath{v}},\chi\dot{\bmath{v}}\rangle\right).
\end{equation}
It follows that
\begin{equation} E[\bmath{v}](t)=E[\bmath{v}](0)=:E_0 \quad {\hbox{ for all }} t\geq 0.
\label{eq:E-con}
\end{equation}
Thus a second stability property holds: the system gives rise to a positive definite energy, which is conserved by the evolution. In particular, no blow-up of the energy is possible.

\subsubsection{Pointwise bounds.}
Conservation of the positive definite energy can be used to show that the perturbation $\bmath{v}$ is bounded throughout the evolution. The argument that follows applies to those background solutions described by Propositions 2 and 3 above. These solutions give rise to functions ${\mathcal{Z}}_{12},{\mathcal{Z}}_{21}, {\mathcal{K}}_{1}$ and ${\mathcal{K}}_{2}$ with the properties required for the arguments below.

It follows from (\ref{eq:E-pos-form}) and (\ref{eq:E-con}) that
\begin{equation}
\| \drs{\bmath{v}}-{\mathcal{Z}}\bmath{v}\|^2 \leq 2E_0,\qquad \langle \bmath{v},{\mathcal{V}}\bmath{v}\rangle \leq 2E_0.
\label{eq:bounds1}
\end{equation}
We note that $E_0$ is the energy of the initial configuration and that we have used (\ref{eq:chi-def}). We can deduce a pointwise stability result using these bounds. Slightly different arguments are required in the soliton and black hole cases.

\paragraph{Solitons.}
In the case of solitonic dyons, we show that (\ref{eq:bounds1}) enables us to obtain an \textit{a priori} bound on the $H^1$ norm of $\bmath{v}$. Then Sobolev embedding leads to a pointwise bound. To see this, we note that
\begin{eqnarray}
\|\bmath{v}\|^2 &=& \int_{r_*=0}^{r_{*{\mathrm {max}}}} v_1^2 + v_2^2 \,dr_* \nonumber \\
&=& \int_{r_*=0}^{r_{*{\mathrm {max}}}} {\mathcal{K}}_1^{-1}{\mathcal{K}}_1 v_1^2 + {\mathcal{K}}_2^{-1}{\mathcal{K}}_2v_2^2 \,dr_*\nonumber\\
&\leq& {\mathcal{C}}_1 \int_{r_*=0}^{r_{*{\mathrm {max}}}} {\mathcal{K}}_1 v_1^2 + {\mathcal{K}}_2v_2^2 \,dr_* \nonumber \\
&=& {\mathcal{C}}_1 \langle\bmath{v},{\mathcal{V}}\bmath{v}\rangle \nonumber \\
&\leq& 2{\mathcal{C}}_1E_0,
\label{eq-v0-bound}
\end{eqnarray}
where
\begin{equation}
{\mathcal{C}}_1 = \max\{ \sup_{r\in[0,\infty)}{\mathcal{K}}_i^{-1}(r), i=1,2\},
\label{eq-ki-max}
\end{equation}
whose existence and positivity is guaranteed by the argument of Section 4.3 above, where we show that the ${\mathcal{K}}_i$ are positive and continuous for $r\in(0,+\infty)$, and satisfy (\ref{eq:ki-soliton-origin}, \ref{eq:ki-soliton-infinity}). Note that the ${\mathcal{K}}_i$ depend only on the background solution functions. Similarly,
\begin{equation}
\| {\mathcal{Z}}\bmath{v}\| \leq {\mathcal{C}}_2 \|\bmath{v}\|,
\label{eq-zv-bound}
\end{equation}
where
\begin{equation}
{\mathcal{C}}_2 = \max\{ \sup_{r\in[0,\infty)}|{\mathcal{Z}}_a(r)|, a=12,21\}.
\label{eq-zij-max}
\end{equation}
Again, the argument of Section 4.3 shows that this term is finite - see (\ref{eq:z12-soliton} -- \ref{z12-z21-infinity}). Then
\begin{eqnarray}
\| \drs\bmath{v}\| &=& \|\drs\bmath{v}-{\mathcal{Z}}\bmath{v}+{\mathcal{Z}}\bmath{v}\| \nonumber \\
&\leq& \|\drs\bmath{v}-{\mathcal{Z}}\bmath{v}\| + \|{\mathcal{Z}}\bmath{v}\| \nonumber \\
&\leq& \sqrt{2E_0} + {\mathcal{C}}_2\sqrt{{\mathcal{C}}_1}\sqrt{2E_0},
\label{eq-v1-bound}
\end{eqnarray}
where we have used (\ref{eq:bounds1}, \ref{eq-v0-bound}, \ref{eq-zv-bound}). Thus we have an \textit{a priori} bound for the $H^1$ norm $\|\bmath{v}\|_1$ of $\bmath{v}$:
\begin{equation}
\|\bmath{v}\|_1^2 := \|\drs\bmath{v}\|^2 + \|\bmath{v}\|^2 \leq 2{\mathcal{C}}_3^2E_0,
\label{eq:h1-bound-soliton}
\end{equation}
where the positive constant ${\mathcal{C}}_3$, determined by ${\mathcal{C}}_1$ and ${\mathcal{C}}_2$, depends only on the background solution. Then the Sobolev inequality \cite{Wald-79}
\begin{equation}
|\bmath{v}(t,r)|^2 \leq \frac12 \left\{ \|\drs\bmath{v}\|^2 + \|\bmath{v}\|^2 \right\}
\label{eq:Sob}
\end{equation}
yields a pointwise bound for the perturbation $\bmath{v}$:
\begin{equation}
|\bmath{v}(t,r)| \leq {\mathcal{C}}_3\sqrt{E_0} \quad {\hbox{ for all }} t\geq 0, r\in[0,+\infty).
\label{eq:solitons-pointwise-stability}
\end{equation}
We note that in these inequalities, $|\cdot|$ is the Euclidean norm in $\mathbb{R}^2$, and so (\ref{eq:solitons-pointwise-stability}) expresses a pointwise bound on the soliton perturbation $\bmath{v}$.

\paragraph{Black holes.}
A slightly different argument is required in the black hole case, due to the different behaviour of the functions ${\mathcal{K}}_i, i=1,2$ at the horizon. The general mathematical argument is the same, relying on the Sobolev inequality to derive a pointwise bound from a bound on the $H^1$ norm. First, we note that
\begin{eqnarray}
\|\drs\bmath{v}\| & \leq & \|\drs\bmath{v}-{\mathcal{Z}}\bmath{v}\| + \|{\mathcal{Z}}\bmath{v}\| \nonumber\\
&\leq & \sqrt{2E_0} + {\mathcal{C}}_4(\langle\bmath{v},{\mathcal{V}}\bmath{v}\rangle)^{1/2}  \nonumber \\
&\leq & (1+{\mathcal{C}}_4)\sqrt{2E_0},
\label{eq:bh-vp-bound}
\end{eqnarray}
where
\begin{equation}
{\mathcal{C}}_4 = \max\left\{ \sup_{r\in[r_h,\infty)}\frac{{\mathcal{Z}}_{21}^2}{{\mathcal{K}}_1}, \sup_{r\in[r_h,\infty)}\frac{{\mathcal{Z}}_{12}^2}{{\mathcal{K}}_2}\right\},
\label{eq:c4}
\end{equation}
which is positive and finite by virtue of the properties listed in (\ref{eq:Ei-hor} -- \ref{eq:k2-inf}). Of particular importance is positivity of the ${\mathcal{K}}_i$, and the finite limits of the relevant ratios at both the horizon $r=r_h$ and at infinity.

Next, we introduce the positive matrix
\begin{equation}
{\mathcal{W}}={\mathcal{V}}^{1/2}=\left( \begin{array}{cc} \sqrt{{\mathcal{K}}_1} & 0 \\ 0 & \sqrt{{\mathcal{K}}_2} \end{array} \right),
\label{eq:Wmat-def}
\end{equation}
and we define
\begin{equation}
\bmath{w} = {\mathcal{W}}\bmath{v}.
\label{eq:wvec-def}
\end{equation}
Notice then
\begin{equation}
\|\bmath{w}\|^2 = \langle\bmath{v},{\mathcal{V}}\bmath{v}\rangle \leq 2E_0. \label{eq:wvec-bound}
\end{equation}
Also,
\begin{eqnarray}
\|\drs\bmath{w}\| & \leq & \|{\mathcal{W}}\drs\bmath{v}\| + \|(\drs{\mathcal{W}})\bmath{v}\| \nonumber \\
& \leq & {\mathcal{C}}_5 \|\drs\bmath{v}\| + {\mathcal{C}}_6 \|\bmath{w}\|,
\label{eq:wvecp-bound}
\end{eqnarray}
where
\begin{equation}
{\mathcal{C}}_5 = \max\left\{ \sup_{r\in[r_h,\infty)}\sqrt{ {\mathcal{K}}_1 }, \sup_{r\in[r_h,\infty)}\sqrt{{\mathcal{K}}_2} \right\},
\label{eq:c5}
\end{equation}
and
\begin{eqnarray}
{\mathcal{C}}_6 &=& \max\left\{ \sup_{r\in[r_h,\infty)}\left|\frac{\drs\sqrt{{\mathcal{K}}_1}}{\sqrt{{\mathcal{K}}_1}}\right|, \sup_{r\in[r_h,\infty)}\left|\frac{\drs\sqrt{{\mathcal{K}}_2}}{\sqrt{{\mathcal{K}}_2}}\right|\right\}
\nonumber \\
&=& \frac12\max\left\{ \sup_{r\in[r_h,\infty)}\left|\mu_0S_0\frac{ {\mathcal{K}}_1'(r) }{ {\mathcal{K}}_1 }\right|, \sup_{r\in[r_h,\infty)}\left|\mu_0S_0\frac{ {\mathcal{K}}_2'(r) }{ {\mathcal{K}}_2 }\right|\right\},
\label{eq:c6}
\end{eqnarray}
both of which are positive and finite by virtue of (\ref{eq:Ei-hor} -- \ref{eq:k2-inf}). Then (\ref{eq:bh-vp-bound}, \ref{eq:wvec-bound}, \ref{eq:wvecp-bound}) establish an \textit{a priori} bound for the $H^1$ norm of $\bmath{w}$, and the Sobolev inequality (\ref{eq:Sob}) applied to $\bmath{w}$ yields
\begin{equation}
|{\mathcal{W}}\bmath{v}| \leq {\mathcal{C}}_7\sqrt{E_0},\label{eq:v-pointwise1}
\end{equation}
where the constant ${\mathcal{C}}_7$ is constructed from ${\mathcal{C}}_4-{\mathcal{C}}_6$, and depends only on the background solution functions. Let us recall certain properties of the ${\mathcal{K}}_i$ established in Section 4.2 above: these functions are positive and continuous on $r\in(r_h,+\infty)$, with finite limits as $r\to+\infty$. They vanish at the horizon, with the asymptotic behaviour ${\mathcal{K}}_i = O(r-r_h)$ as $r\to r_h$. It follows from (\ref{eq:Wmat-def}) and (\ref{eq:v-pointwise1}) that $\bmath{v}$ is bounded on any interval of the form $[r_h+\epsilon,+\infty)$ with $\epsilon>0$. The boundary condition $\lim_{r\to r_h}\bmath{v}=0$ ensures that this pointwise bound extends to the entire interval $[r_h,+\infty)$. This establishes a pointwise bound for $\bmath{v}(t,r)$ for all $t\geq 0$.

\section{Conclusions}
\label{sec:conc}

In this paper we have proven the existence of dyonic soliton and black hole solutions of four-dimensional ${\mathfrak {su}}(2)$ Einstein-Yang-Mills theory in asymptotically anti-de Sitter space which are stable under linear, spherically symmetric, perturbations of the metric and gauge field.
Although the static, dyonic, equilibrium solutions of the field equations were found numerically over fifteen years ago \cite{Bjoraker}, their stability has not been investigated until now.

The perturbation equations for linear, spherically symmetric, perturbations of the dyonic equilibrium solutions are much more complicated than those for purely magnetic equilibrium solutions, which is perhaps why the stability question has not been explored previously in the literature.  In the purely magnetic case, with a suitable choice of Lie algebra gauge, the perturbation equations decouple into two sectors, known as the ``sphaleronic'' (odd-parity) and ``gravitational'' (even-parity) sectors \cite{Lavrelashvili1}. This decoupling of the equations for odd- and even-parity perturbations greatly simplifies the analysis (see, for example, \cite{Straumann} for the ${\mathfrak {su}}(2)$ case and \cite{Baxter} for the larger ${\mathfrak {su}}(N)$ gauge group).
In the purely magnetic case this decoupling arises because the static equilibrium solutions are both spherically symmetric and invariant under a parity reflection ${\bmath {x}}\rightarrow -{\bmath {x}}$ of the space co-ordinates and the purely magnetic gauge field.
Therefore the perturbations which are odd and even under the parity reflection can be considered separately.
While the dyonic static solutions considered here are still spherically symmetric, the gauge field is not invariant under a parity transformation (see, for example, the discussion in section 2 of \cite{Volkov1}) and so the perturbation equations for the odd- and even-parity perturbations no longer decouple.

We have therefore taken an alternative approach in this paper, and considered perturbations which are invariant under Lie algebra gauge transformations.
The perturbations of the metric functions can be found in terms of the gauge field perturbations and hence eliminated.
After some manipulation, the equations for the remaining three gauge field perturbations can be cast into a pair of coupled Schr\"odinger-like equations (\ref{eq:eigenvalue}) involving just two of the perturbations ($\delta \kappa$ and $\delta \omega $) and a single constraint equation (\ref{eq:deltaxifinal}) which does not contain any time derivatives and determines the third perturbation ($\delta \xi $) once the other two are known.
The lack of a dynamical equation for $\delta \xi $ is due to a remaining diffeomorphism gauge freedom, corresponding to a redefinition of the time co-ordinate.

In \cite{Nolan} we proved the existence of static, spherically symmetric, dyonic soliton and black hole solutions of the ${\mathfrak {su}}(2)$ EYM equations
in adS for which the single magnetic gauge field function $\omega _{0}$ has no zeros.  These nodeless solutions exist for any value of the negative cosmological constant $\Lambda <0$, in a neighbourhood of the trivial (Schwarzschild-adS) embedded solution.
By analysing the afore-mentioned pair of coupled Schr\"odinger-like perturbation equations, we have been able to prove that nodeless dyonic solutions sufficiently close to the embedded trivial solution are stable under linear, spherically symmetric, perturbations of the metric and gauge field.
This extends the proof of the existence of stable purely magnetic solutions \cite{Bjoraker,Winstanley1} to the dyonic case.

It would be interesting the explore the consequences of the existence of stable dyonic black holes for the ``no-hair'' conjecture, in the form stated by Bizon \cite{Bizon2}, namely
\begin{quotation}
Within a given matter model, a {\em {stable}} stationary black hole is uniquely
determined by global charges.
\end{quotation}
Combining the results of \cite{Baxter,Shepherd}, we have evidence that the above conjecture is true for purely magnetic black holes in ${\mathfrak {su}}(N)$ EYM in adS.  To investigate whether the stable dyonic black holes, whose existence we have proven here, satisfy the above conjecture, appropriate electric and magnetic charges would need to be defined, and then one would need to determine whether these charges uniquely characterize the black holes.
We hope that this question will be the subject of further research.

Dyonic black hole solutions of EYM in adS have received a great deal of attention recently in the literature when the event horizon, rather than being topologically spherical  as considered in this paper, has planar topology.
In \cite{Gubser} it was found that there is a second order phase transition from the embedded planar Reissner-Nordstr\"om-adS solution to a black hole with a
non-trivial dyonic YM condensate.
There is now a substantial literature on such planar dyonic EYM black holes as models of $p$-wave holographic superconductors (see, for example, \cite{holographic} for a selection of references, and \cite{holreview} for a recent review).
The thermodynamic behaviour of these planar dyonic EYM black holes has been studied extensively (this is the key aspect of their interpretation as holographic superconductors) but little is known about their classical stability.
At least some purely magnetic topological black holes in ${\mathfrak {su}}(2)$ \cite{Radu} and ${\mathfrak {su}}(N)$ \cite{Baxter2} EYM in adS
have been proven to be stable when the magnetic gauge field functions have no zeros.
In this paper we have considered the stability of spherically symmetric dyonic solutions of ${\mathfrak {su}}(2)$ EYM, and a natural question would be to extend this to topological black holes, or to a larger gauge group (the existence of dyonic solitons and black holes in ${\mathfrak {su}}(N)$ EYM in adS having recently been established \cite{Baxter3}).
However, our results in this paper are for dyonic solutions where the magnetic gauge field function $\omega _{0}$ has no zeros, whereas for the solutions of relevance for holographic superconductors $\omega _{0}$ has a single zero, located on the adS boundary.
This will complicate the classical stability analysis and therefore we leave this as an open question for future investigation.

\ack
We acknowledge support from the Office of the Vice President for Research in Dublin City University for an International Visitor Programme grant which enabled the completion of this work.
The work of E.W. is supported by the Lancaster-Manchester-Sheffield
Consortium for Fundamental Physics under STFC grant ST/L000520/1.
E.W. thanks Dublin City University for hospitality while this work was in progress.


\section*{References}

\begin{thebibliography}{99}

\bibitem{Bartnik}
Bartnik R and McKinnon J 1988 \PRL {\bf {61}} 141--4

\bibitem{AFBHs}
Volkov M S and Gal'tsov D V 1989 {\it {JETP Lett}} B {\bf {50}} 346

\nonum
Volkov M S and Gal'tsov D V 1990 {\it {Sov.~J.~Nucl.~Phys.}} {\bf {51}} 747

\nonum
Bizon P 1990 \PRL {\bf {64}} 2844--7

\nonum
Kunzle H P and Masood-ul-Alam A K M 1990 \JMP {\bf {31}} 928--35

\bibitem{Volkov1}
Volkov M S and Gal'tsov D V 1999 {\it {Phys.~Rept.}} {\bf {319}} 1--83

\bibitem{Ershov}
Ershov A A and Gal'tsov D V 1989 \PL A {\bf {138}} 160--4

\nonum
Ershov A A and Gal'tsov D V 1990 \PL A {\bf {150}}  159--62

\bibitem{Straumann}
Straumann N and Zhou Z 1990 \PL B {\bf {237}} 353--6

\nonum
Straumann N and Zhou Z 1990 \PL B {\bf {243}} 33--5

\nonum
Gal'tsov D V and Volkov M S 1992 \PL A {\bf {162}} 144--8

\nonum
Volkov M S and Gal'tsov D V 1995 \PL B {\bf {341}} 279--85

\nonum
Hod S 2008 \PL B {\bf {661}} 175--8

\bibitem{Bjoraker}
Bjoraker J and Hosotani Y 2000 \PRL {\bf {84}} 1853--6

\nonum
Bjoraker J and Hosotani Y 2000 \PR D {\bf {62}} 043513

\bibitem{Winstanley1}
Winstanley E 1999 \CQG {\bf {16}} 1963--78

\bibitem{Sarbach}
Sarbach O and Winstanley E 2001 \CQG {\bf {18}} 2125--46

\nonum
Winstanley E and Sarbach O 2002 \CQG {\bf {19}} 689--724

\bibitem{Baxter1}
Baxter J E, Helbling M and Winstanley E 2007 \PR D {\bf {76}} 104017

\nonum
Baxter J E, Helbling M and Winstanley E 2008 \PRL {\bf {100}} 011301

\nonum
Baxter J E and Winstanley E 2008 \CQG {\bf {25}} 245014

\bibitem{Baxter}
Baxter J E and Winstanley E 2015
On the stability of soliton and hairy black hole solutions of ${\mathfrak {su}}(N)$ Einstein-Yang-Mills theory with a negative cosmological constant
arXiv:1501.07541

\bibitem{Nolan}
Nolan B C and Winstanley E 2012 \CQG {\bf {29}} 235024

\bibitem{Kunzle}
Kunzle H P 1991 \CQG {\bf {8}} 2283--97

\bibitem{Winstanley}
Winstanley E 2009 {\it {Lect.~Notes Phys.}} {\bf {769}} 49--87

\bibitem{Shepherd1}
Shepherd B L and Winstanley E 2015 Dyons and dyonic black holes in ${\mathfrak {su}}(N)$ Einstein-Yang-Mills theory in anti-de Sitter arXiv:1512.03010

\bibitem{Gerlach-Sengupta}
Gerlach U H and Sengupta U K 1979 \PR D {\bf {19}} 2268--2272

\bibitem{Wald-79}
Wald R M 1979 \JMP {\bf {20}} 1056--1058

\nonum
Wald R M 1980 \JMP {\bf {21}} 218

\bibitem{Lavrelashvili1}
Lavrelashvili G V and Maison D 1995 \PL B {\bf {343}} 214--7

\bibitem{Bizon2}
Bizon P 1994 {\it {Acta Phys.~Polon.}} B {\bf {25}} 877--98

\bibitem{Shepherd}
Shepherd B L and Winstanley E 2012 \CQG {\bf {29}} 155004

\bibitem{Gubser}
Gubser S S 2008 \PRL {\bf {101}} 191601

\bibitem{holographic}
Gubser S S and Pufu S S 2008 {\it {J.~High Energy Phys.}} {\bf {0811}} 033

\nonum
Manvelyan R, Radu E and Tchrakian D H 2009 \PL B {\bf {677}} 79--87

\nonum
Herzog C P and Pufu S S 2009 {\it {J.~High Energy Phys.}} {\bf {0904}} 126

\nonum
Peeters K, Powell J and Zamaklar M 2009 {\it {J.~High Energy Phys.}} {\bf {0909}} 101

\nonum
Gubser S S, Rocha F D and Yarom A 2010 {\it {J.~High Energy Phys.}} {\bf {1011}} 085

\nonum
Ammon M, Erdmenger J, Grass V, Kerner P and O'Bannon A 2010 \PL B {\bf {686}} 192--8

\nonum
Akhavan A and Alishahiha M 2011 \PR D {\bf {83}} 086003

\nonum
Gangopadhyay S and Roychowdhury D 2012 {\it {J. High Energy Phys.}} {\bf {1208}} 104

\nonum
Arias R E and Landea I S 2013 {\it {J. High Energy Phys.}} {\bf {1301}} 157

\nonum
Herzog C P, Huang K-W, Vaz R 2014 {\it {J.~High Energy Phys.}} {\bf {1411}} 066

\nonum
Devecio\u{g}lu D O 2014 \PR D {\bf {89}} 124020

\nonum
Arean D, Farahi A, Pando Zayas L A, Landea I S and Scardicchio A 2015 {\it {J.~High Energy Phys.}} {\bf {1507}} 046

\nonum
Nie Z-Y,  Cai R-G, Gao X, Li L and Zeng H 2015 {\it {Eur.~Phys.~J.}} C {\bf {75}} 559

\nonum
Fan Z Y and Lu H 2015 \PL B {\bf {743}} 290--4

\bibitem{holreview}
Cai R G, Li L, Li L F and Yang R Q 2015
  {\it {Sci. China Phys.~Mech.~Astron. }} {\bf {58}} 060401

\bibitem{Radu}
van der Bij J and Radu E 2002 \PL B {\bf {536}} 107--13

\bibitem{Baxter2}
Baxter J E 2015 Stable topological hairy black holes in ${\mathfrak {su}}(N)$ EYM theory with $\Lambda <0$ arXiv:1507.03127

\bibitem{Baxter3}
Baxter J E 2015 Existence of topological hairy dyons and dyonic black holes in anti-de Sitter ${\mathfrak {su}}(N)$ EYM theory arXiv:1507.05314

\end{thebibliography}
\end{document}